\documentclass[twocolumn,prd,nofootinbib,superscriptaddress]{revtex4}
\newcommand\ForInternalReference[1]{}
\newcommand\SkipForEarlyCirculation[1]{#1}
\newcommand\SkipBoxing[1]{}
\usepackage{verbatim}
\usepackage{graphicx}
\usepackage{dcolumn}
\usepackage{bm}
\usepackage{color}
\usepackage{xspace}
\usepackage{url}
\usepackage{amsmath}
\usepackage{float}
\usepackage{multirow}
\usepackage{times}
\newcommand\optional[1]{}

\newcommand\E[1]{\left\langle #1\right\rangle}

\newcommand\lnL{ \ln {\cal L}}
\newcommand\lnLmarg{ \ln{\cal L}_{\rm marg}{}}
\newcommand\unit[1]{{\rm #1}}

\newcommand\ILE{ILE}
\newcommand\editremark[1]{{\color{red} #1}}
\usepackage{color}
\definecolor{amber}{rgb}{1.0, 0.75, 0.0}
\definecolor{orange}{rgb}{1.0, 0.5, 0.0}
\definecolor{amaranth}{rgb}{0.9, 0.17, 0.31}

\graphicspath{{./figures/}}
\newcommand{\mc}{{\cal M}}

\newcommand\LambdaTilde{\widetilde{\Lambda}}
\newcommand\DeltaLambdaTilde{\delta \widetilde{\Lambda}}
\def\ltsima{$\; \buildrel < \over \sim \;$}
\def\simlt{\lower.5ex\hbox{\ltsima}}
\def\gtsima{$\; \buildrel > \over \sim \;$}
\def\simgt{\lower.5ex\hbox{\gtsima}}

\newcommand{\IMRPD}{\textsc{IMRPhenomD}\xspace}
\newcommand{\IMRPDT}{\textsc{IMRPhenomD\_NRTidal}\xspace}
\newcommand{\IMRP}{\textsc{IMRPhenomPv2}\xspace}
\newcommand{\SEOBP}{\textsc{SEOBNRv3}\xspace}
\newcommand{\SEOBA}{\textsc{SEOBNRv4}\xspace}
\newcommand{\SEOBAROM}{\textsc{SEOBNRv4\_ROM}\xspace}
\newcommand{\NRSur}{NRSur7dq2\xspace}
\newcommand{\TEOB}{SEOBNRv4T\xspace}
\newcommand{\Resum}{TEOBResumS\xspace}
\newcommand\RIFT{RIFT}

\def\RIT{Center for Computational Relativity and Gravitation, Rochester Institute of Technology, Rochester, New York 14623, USA}

\begin{document}

\title{Rapid and accurate parameter inference for coalescing, precessing compact binaries}
\author{J. Lange}
\affiliation{\RIT}
\author{R. O'Shaughnessy}
\affiliation{\RIT}
\author{M. Rizzo}
\affiliation{\RIT}

\begin{abstract}
Extending prior work by Pankow et al, we introduce RIFT, an algorithm to  perform Rapid parameter inference on gravitational wave sources via Iterative
FiTting.  
We demonstrate this approach can correctly recover the parameters of coalescing compact binary systems, using detailed
comparisons of RIFT to the well-tested LALInference software library.
We provide several examples where the unique speed and flexibility of RIFT enables otherwise intractable or awkward
parameter inference analyses, including (a) adopting  either  costly and novel models for outgoing gravitational waves;
and (b)  mixed  approximations, each suitable to different parts  of the compact binary parameter space.  
We demonstrate how \RIFT{} can be applied to binary neutron stars, both for parameter inference and direct constraints
on the nuclear equation of state.
\end{abstract}
\maketitle

\section{Introduction}
The Advanced LIGO  \cite{2015CQGra..32g4001T}  and Virgo \cite{gw-detectors-Virgo-original-preferred}  ground-based gravitational wave (GW) detectors have
identified several coalescing compact binaries \cite{DiscoveryPaper,LIGO-O1-BBH,2017PhRvL.118v1101A,LIGO-GW170814,LIGO-GW170608,LIGO-GW170817-bns}.  Over the next few years, hundreds more will be
confidently identified, with proportionally more uncertain candidates at the margins of detector sensitivity.  These
sources provide an opportunity and a challenge.  On the one hand, reliable, unbiased  parameter inferences will be
essential to achieve the full potential of high-precision tests of general relativity; constraints on nuclear matter;
and inferences about present-day populations and progenitor astrophysics.    On the other hand, these inferences will be
stable only with a careful exploration of model systematics, which at present cannot be performed due to their
computational cost.  For example, using the standard parameter estimation pipeline LALInference (LI) \cite{gw-astro-PE-lalinference-v1}, only
special-purpose approximations and infrastructure can enable generic analyses with computational costs measured in less than days (for binary
black holes) or weeks (for binary neutron stars) and simultaneously produce production-ready results which
comprehensively estimate all binary parameters.

To address this challenge, Pankow et al. \cite{gwastro-PE-AlternativeArchitectures}  proposed an  alternative and highly-parallelizable grid-based parameter
estimation strategy.   In this approach,  each candidate GW signal is compared  to a
regular grid of candidate source parameters, producing an array of candidate likelihood values, a process
henceforth denoted by \ILE.      Interpolating the likelihood over the grid, Pankow et al
\cite{gwastro-PE-AlternativeArchitectures}  estimated the masses of low-mass compact objects.    
Since the original study, several investigations have applied a similar  method to other masses and higher
dimension \cite{NRPaper,2017PhRvD..96j4041L,2017CQGra..34n4002O}.  Due to the exponentially-increasing number of points
appearing in a regular grid in high dimension, these later approaches usually adopted a different approach for analyzing
strong signals, approximating the likelihood by a Gaussian in suitable coordinates.  

In this work, building on early changes adopted in  \cite{2017CQGra..34n4002O}, we introduce and demonstrate a
straightforward  generalization of the approach in Pankow et al.   To circumvent the cost and limitations of polynomial interpolation and periodic grids, we perform  gaussian
process interpolation on an unstructured grid.  Further,
to create and validate the grid used for interpolation, we employ an iterative procedure, using estimates of the
posterior distribution at stage $k$ to propose new grid points to augment generation $k+1$.  
We demonstrate this procedure can rapidly and accurately recover the parameters of generic quasicircular binaries, including
models and scenarios whose computational cost would be prohibitive for LI.  
This approach has been developed and applied to GW observations %
since 2016,  particularly in the context of comparisons to numerical relativity.  

This paper is organized as follows.  
In Section  \ref{sec:Methods} we describe our parameter estimation pipeline, including  the underlying gaussian process interpolation;
iterative evaluation of ILE on the posterior grid; and convergence tests.  
In Section \ref{sec:Models} we enumerate the concrete models used for parameter inference in this work.  
In Section \ref{sec:Validation} we describe several validation studies we performed to assess it, including purely
synthetic likelihoods and head-to-head comparison with LI.
In Section \ref{sec:Applications} we describe applications of our pipeline, particularly to assessing the impact of
model systematics.  %
We summarize our results and discuss future generalizations in Section \ref{sec:conclude}.
In a  companion paper \cite{gwastro-PENR-FitHybridPlacement}, we  demonstrate how the likelihood fitting techniques
developed in this work can be applied to direct comparison of GW measurements with solutions of Einstein's equations, as
in \cite{NRPaper,2017PhRvD..96j4041L}.

\section{Parameter inference via iterative gaussian process fits}
\label{sec:Methods}

\subsection{Coordinates and notation}

A coalescing compact binary in a quasicircular orbit can be completely characterized by its intrinsic
parameters, namely its individual masses $m_i$, spins $\mathbf{S}_i$, and by internal multipole moments which
characterize its matter degrees of freedom like the individual dimensionless tidal deformabilities $\Lambda_i$ (see
Appendix \ref{subap:tides}) ; and its seven extrinsic parameters: right
ascension, declination, luminosity distance, coalescence time, and three Euler angles characterizing its orientation
(e.g., inclination, orbital phase, and polarization).  In this work, we will also use the total mass $M=m_1+m_2$ and mass ratio q defined in the following way:
\begin{equation}
\label{eq:q}
q=m_{1}/m_{2},\textrm{where} \quad m_1>m_2.
\end{equation}
With regard to spin, we define an effective spin \cite{2001PhRvD..64l4013D,2008PhRvD..78d4021R,2011PhRvL.106x1101A}, which is a combination of the spin components along to orbital angular momentum, in the following way,
\begin{equation}
\label{eq:chieff}
\chi_{\rm eff}=(\bm{S_{1}}/m_{1}+\bm{S_{2}}/m_{2})\cdot\hat{L}/M
\end{equation}
where $\bm{S_{1}}$ and $\bm{S_{2}}$ are the dimensionless spins on the individual BH defined:
\begin{equation}
\label{eq:chii}
\bm{\chi_{i}}=\bm{S_{i}}/m_{i}^{2}.
\end{equation}
We will express the dimensionless spins in terms of cartesian components $\chi_{i,x},\chi_{i,y}, \chi_{i,z}$, expressed
relative to a frame with $\hat{z}=\hat{L}$ and (for simplicity) at the orbital frequency corresponding to the earliest
time of astrophysical interest (e.g., an orbital frequency of $\simeq 10 \unit{Hz}$).

\subsection{Marginalized likelihoods}
\ILE{} provides a straightforward and efficient mechanism to compare any specific candidate gravitational wave source with
real or synthetic data   \cite{gwastro-PE-AlternativeArchitectures,NRPaper,2017PhRvD..96j4041L,2017CQGra..34n4002O},
by marginalizing the likelihood of the data over the seven coordinates characterizing the binary's coalescence event and
orientation relative to the earth.  
Specifically the likelihood of the data given gaussian noise has the form  (up to normalization)
\begin{equation}
\label{eq:lnL}
\ln {\cal L}(\bm{\lambda} ;\theta )=-\frac{1}{2}\sum\limits_{k}\langle h_{k}(\bm{\lambda} ,\theta )-d_{k} |h_{k}(\bm{\lambda} ,\theta )-d_{k}\rangle _{k}-\langle d_{k}|d_{k}\rangle _{k},
\end{equation}
where $h_{k}$ are the predicted response of the k$^{th}$ detector due to a source with parameters ($\bm{\lambda}$, $\theta$) and
$d_{k}$ are the detector data in each instrument k; $\bm{\lambda}$ denotes the combination of redshifted mass $M_{z}$ and the
remaining parameters needed to uniquely specify the binary's dynamics; $\theta$ represents the
seven extrinsic parameters (4 spacetime coordinates for the coalescence event and 3 Euler angles for the binary's
orientation relative to the Earth); and $\langle a|b\rangle_{k}\equiv
\int_{-\infty}^{\infty}2df\tilde{a}(f)^{*}\tilde{b}(f)/S_{h,k}(|f|)$ is an inner product implied by the k$^{th}$ detector's
noise power spectrum $S_{h,k}(f)$. 
In practice we adopt a low-frequency cutoff f$_{\rm min}$ so all inner products are modified to
\begin{equation}
\label{eq:overlap}
\langle a|b\rangle_{k}\equiv 2 \int_{|f|>f_{\rm min}}df\frac{\tilde{a}(f)^{*}\tilde{b}(f)}{S_{h,k}(|f|)}.
\end{equation}
The joint posterior probability of $\bm{\lambda} ,\theta$ follows from Bayes' theorem:
\begin{equation}
p_{\rm post}(\bm{\lambda} ,\theta)=\frac{ {\cal L}(\bm{\lambda} ,\theta)p(\theta)p(\bm{\lambda})}{\int d\bm{\lambda} d\theta {\cal L}(\bm{\lambda} ,\theta)p(\bm{\lambda})p(\theta)},
\end{equation}
where $p(\theta)$ and $p(\bm{\lambda})$ are priors on the (independent) variables $\theta ,\bm{\lambda}$. For each $\bm{\lambda}$, we evaluate the marginalized likelihood
\begin{equation}
 {\cal L}_{\rm marg}\equiv\int  {\cal L}(\bm{\lambda} ,\theta )p(\theta )d\theta
\end{equation}
via direct Monte Carlo integration, where $p(\theta)$ is uniform in 4-volume and source orientation.  
To evaluate the likelihood in regions of high importance, we use an adaptive Monte Carlo as described in
\cite{gwastro-PE-AlternativeArchitectures}.    As described in Pankow et al, this marginalized likelihood can be evaluated efficiently
because, having generated the dynamics and outgoing radiation in all possible directions once and for all for fixed
$\mathbf{\lambda}$, the likelihood can be evaluated as a function of $\theta$ at very low computational cost.

Using Bayes' theorem, the posterior distribution for intrinsic parameters $\bm{\lambda}$ can be expressed as 
\begin{equation}
\label{eq:post}
p_{\rm post}=\frac{{\cal L}_{marg}(\bm{\lambda} )p(\bm{\lambda})}{\int d\bm{\lambda} {\cal L}_{\rm marg}(\bm{\lambda} ) p(\bm{\lambda} )}.
\end{equation}
where prior $p(\bm{\lambda})$ is the prior on intrinsic parameters like mass and spin.  
ILE itself will only provide point estimates $\lnLmarg_{\alpha}$ given proposed evaluation points $\bm{\lambda}_\alpha$, not
the interpolated $\lnLmarg$ necessary to construct a full solution.  

\subsection{Gaussian process interpolation}
Given some proposed training data  $\{(\bm{\lambda}_\alpha,\lnLmarg_\alpha)\}$, we estimate $\lnLmarg$ via conventional Gaussian Process interpolation
\cite{book-Rasmussen-GP} to produce a weakly nonparametric interpolation and error estimate.   Gaussian process interpolation has already been used in the field
\cite{2014PhRvD..90j4012V,2014PhRvL.113y1101M,2017CQGra..34n4002O,2017PhRvD..96l3011D,2016PhRvD..93f4001M,2017CQGra..34n4002O}, particularly to
propagate uncertainties.
In this approach, we estimate the expected value of  $y(x)$ from data $x_*$ and values $y_*$ via
\begin{eqnarray}
\E{y(x)} = \sum_{\alpha,\alpha'} k(x,x_{*,\alpha}) (K^{-1})_{\alpha,\alpha'} y_{*,\alpha'}
\end{eqnarray}
where $\alpha $ is an integer running over the number of training samples in $(x_*,y_*)$ and where the matrix  $K =
k(x_\alpha,x_\alpha')$ $y_*$.   [For brevity and to be consistent with conventional notation, in this section we denote
  $\bm{\lambda}_\alpha$ by $x$ and $\lnLmarg_\alpha$ by $y$.]   We employ a kernel function $k(x,x')$ which allows  for uncertainty in each
estimated training point's value $y_{*,\alpha}$ due to Monte Carlo integration, as well as a conventional squared
exponential kernel to allow for changes in the functions versus parameters:
\begin{eqnarray}
k(x,x') =  \sigma_o^2 e^{-(x-x')Q(x-x')/2} + \sigma_n^2 \delta_{x,x'}
\end{eqnarray}
Following usual practice and to insure flexibility for generic sources, we optimize the hyper-parameters
$\sigma_o,\sigma_n$ and the positive-definite symmetric matrix $Q$ on our training data.
We perform all gaussian process interpolation with widely-available open-source software \cite{scikit-learn}.  
The computational cost of full-scale Gaussian process optimization and evaluation increases rapidly with the dimension $D$ of the
matrix $K$, as $D^3$ and $D^2$ respectively; for typical hardware, we are presently limited to $O(10^4)$ training points in $8$
dimensions.   In future work we will employ other techniques like sparse approximations and multilayer
(``pool-of-experts'') designs, which can achieve comparable results with better scaling. %

\subsection{Flexible Monte Carlo generation of posterior samples}

We use a single code (henceforth denoted by CIP) to Construct the Intrinsic Posterior distribution, by  loading training data
$\{(\bm{\lambda}_\alpha,\lnLmarg_\alpha)\}$;  fitting that data, using some coordinate system $X$ for $\{\bm{\lambda}\}$; and then constructing
samples from the  posterior distribution via adaptive Monte Carlo using another coordinate system $Y$ with fiducial prior
distributions $p(y)$, employing the same adaptive
Monte Carlo techniques described \cite{gwastro-PE-AlternativeArchitectures} and applied in ILE.  
When fitting the likelihood, we employ coordinate systems well-adapted to the likelihood, which based on many decades of
theoretical and computational studies are likely to produce an approximately gaussian likelihood in the limit of strong
signals (e.g., coordinates in which the Fisher matrix is a good approximation to the log-likelihood in the high-SNR
limit).   When performing a Monte Carlo integral to construct a posterior distribution, however, we employ coordinates $Y$
that allow us to efficiently specify the priors in separable form and simultaneously sample the parameter space
thoroughly in the region with significant prior support.  
For example, we often use the chirp mass  $\mc=(m_1 m_2)^{3/5}/(m_1+m_2)^{1/5}$ as a sampling coordinate, to insure our
adaptive Monte Carlo method can efficiently identify the often exceptionally narrow region of $\mc$ consistent with the
posterior, particularly for binary neutron stars.   
In Appendix \ref{ap:CoordinatesAndPriors} we describe the specific array of coordinate systems and priors we use.

As with \ILE, the adaptive Monte Carlo produces a sequence of weighted points $w_k,\bm{\lambda}_k$ that fully
characterize the posterior distribution.  For example, any
one-dimensional marginal distribution for a function $g(\bm{\lambda})$ can be computed via the corresponding weighted sum
$P(<G) = [\sum_k w_k \Theta(g(\bm{\lambda}_k) -G)]/\sum_q w_q$, where $\Theta$ is the Heavyside function.  To improve compatibility with other codes' output and to minimize overhead -- in
practice, a set of weighted posterior samples often includes $ \gg 10^8$ points, many with low weight -- we then
uniformly resample from the weighted posterior points, with replacement.  Specifically, after ordering the sequence
$w_k$ so $w_1 \le w_2 \le \ldots$, we draw $N$ uniformly
distributed random numbers $u_q$ and choose $\bm{\lambda}_q$ such that $q$ is the largest number with $\sum_{k\le q} w_k
/\sum_k w_k  <  u_q$.
Our adaptive Monte Carlo procedure currently requires separable priors, as described in Appendix
\ref{ap:CoordinatesAndPriors}.   We generate results for generic priors by  reweighting the underlying weighted samples
produced above, before performing the draw-with-replacement procedure to generate a fair sample from the target posterior.  For example, if $\{(w_k,\bm{\lambda}_k)\}$ are generated by the procedure above with a prior $p_{\rm
  ref}(\bm{\lambda})$, then we generate posterior distributions suitable to a prior $p(\bm{\lambda})$ via reweighted samples
$\{(w_k p(\bm{\lambda}_k)/p_{\rm ref}(\bm{\lambda}_k),\bm{\lambda}_k )\}$, following the procedures outlined above.

\subsection{Iterative procedure}
The fitting and sampling procedures described above produces a proposed set of posterior samples $B_0=\{\bm{\lambda}_k\}$, given training data
$A_0=\{\bm{\lambda}_\alpha\}$.   We then use \ILE{} to evaluate $\lnLmarg$ on $\bm{\lambda}_k$, then
  perform the procedure described above starting with revised training data $A_1 = A_0 \cup B_0$ to produce a new set of
proposed samples $B_1$.  By repeating this process several times, we can  validate and refine our fit and hence
posterior.  
To assess whether the fit and posterior have converged, we use (all of) the one-dimensional marginal distributions,
comparing the empirical cumulative distributions in one iteration against the next using standard tools; see Appendix \ref{ap:validate:Gaussians}.
As described in  Appendix \ref{ap:validate:Gaussians}, we  tested this iterative fitting and posterior generation procedure using
synthetic likelihood functions $\lnLmarg$ for which the posterior distributions can be calculated analytically.   This
procedure succeeds consistently for all dimensions we thoroughly tested ($d \le 6$).

We  extended the basic framework described above to make it more robust and efficient.  
To make the process more robust, we  employ a dithering step.  Specifically, given (a subset of) $B_k$ and its covariance $\Sigma$ in the subset of
parameters we choose to dither,  we can add random offsets $\Delta \bm{\lambda}_k$ drawn from a multinormal distribution with
covariance $\epsilon^2\Sigma$, where $\epsilon$ is a factor controlling the amount of dithering.  After removing samples
which due to $\Delta \bm{\lambda}_k$ are no longer physical, we have a new set $C_k$.  The iterative procedure uses both undithered and dithered
points, so  $A'_{k+1}=A'_k\cup B_k \cup C_k$.  We typically employ a dithering factor $\epsilon \simeq 1$.
Second, to make the process more efficient, we can (if needed) perform the first several iterations with fewer and
physically-motivated degrees of freedom: the map $X(\bm{\lambda})$ used to perform the fit will have fewer dimensions than
$\bm{\lambda}$.   This approach may be required if the initial training set $A_0$ is too small to provide a useful fit
in all dimensions of $\bm{\lambda}$.   If the likelihood can be approximated at some level using only the variables in $X$ -- if only that subset
of variables characterizes the most easily measurable quantities  --   then these dimensionally-reduced iterations cause the remaining
degrees of freedom to be efficiently and randomly explored.   As a concrete example,
for  massive binary black holes like GW150914, even if precessing, the likelihood can be well approximated using just $\mc,\eta,
\chi_{eff}$ \cite{NRPaper}, though after more iterations our fitting procedure  captures more  subtle features of the
fit and hence posterior.
After these first few seed iterations, we revert back to the conventional approach described above.

Gaussian processes are expensive, with evaluation cost scaling as the number of points squared.   Before fitting, we always eliminated
points whose marginalized likelihoods were many orders of magnitude outside the expected support of the
distribution (i.e., with $\ln {\cal L} \lesssim \ln {\cal L}_{\rm max} - F \chi^2_{d}(0.9)/2$ where $\chi^2_d(x)$ is the
inverse-chisquared distribution with $d$ degrees of freedom, and $F\simeq 10$).  
In rare cases, even after this condition is applied, we  iteratively accumulate a number of points $N$ greater than our
computationally-tractable limit $N_*$.  In these cases, we   randomly subdivide our training points into $G$ equal-sized
subgroups smaller than $N_*$.    In one approach, we  repeat our analysis on each random subsample, constructing
$b=1\ldots G$  
gaussian-process approximations  $g_b(\bm{\lambda})$ and posterior samples $A_b$, requiring consistency between these outputs $A_b$.  In another approach,
we construct the posterior distribution using the average $\bar{g}(\bm{\lambda}) \equiv \sum_b g(\bm{\lambda})/G$.

As in previous work \cite{gwastro-PE-AlternativeArchitectures,NRPaper}, we analyze data containing candidate signals
starting with a good approximation to (some of) that sources' parameters, provided by the search algorithm which flagged
this stretch of data as a candidate binary coalescence.  We use this information to conservatively identify the initial
grid $A_o$ of test parameters to explore; see, e.g., \cite{gwastro-PE-AlternativeArchitectures} and references therein.

\subsection{Reconstructing source-frame binary masses}

\ILE{}  compares waveforms with fixed  detector-frame masses $m_{i,z}=m_i(1+z)$ to real and synthetic observational
data.  As a result, the procedures described above (and in all previous work) produce posterior distributions as a
function of $M_z=(m_1+m_2)(1+z)$ and dimensionless intrinsic variables.    We must perform some additional
post-processing and analysis to recover the distribution of the total source-frame mass
$M=(m_1+m_2)$  and its joint distribution with all other intrinsic parameters.

The most straightforward and robust procedure to reconstruct the full, joint source-frame posterior reprocesses
 the output of \ILE{} from the final set of draws $A_N$ from the iterative procedure described above.  Each
point $\bm{\lambda}_k \in A_N$ is a fair draw from the posterior distribution, by design.   In the process of performing the
Monte Carlo integral for $\lnLmarg$ for each $\bm{\lambda}_k$, 
\ILE{} produces  weighted posterior samples $\{w_{k,\alpha},\theta_{k,\alpha}\}$, where the parameters
$\theta_{k,\alpha}$ include the distance and therefore redshift $z$.  We generate the joint posterior distribution on
source-frame parameters by  combining all \ILE{} output samples and re-expressing all masses using a suitable redshift.  For example, we can
generate the cumulative distribution of $M$ via $P(<M) = \sum_{k,\alpha} w_{k,\alpha} \Theta(M - M_{z,k}/(1+z_{k,\alpha}))$.

\section{Models and sources in this work }
\label{sec:Models}

In this work, we perform parameter inference with several standard approximations to the outgoing radiation of
coalescing binaries, which fall in three families:  effective one body (EOB) models
\cite{2014PhRvD..89f1502T,2014PhRvD..89h4006P}, specifically 
\SEOBP,
\SEOBA, and 
\SEOBAROM; 
phenomenological frequency-domain inspiral and merger models,
specifically using \IMRPD
and \IMRP~\cite{gwastro-mergers-IMRPhenomP}; and surrogate waveforms, directly calibrated to numerical relativity
\cite{2015PhRvL.115l1102B,2017PhRvD..95j4023B,2017PhRvD..96b4058B}, specifically using \NRSur.

The EOB approach models the inspiral and spin dynamics
of coalescing binaries via an ansatz for the two-body Hamiltonian~\cite{gw-astro-EOBspin-Tarrachini2012}, which is
solved in the time domain. 
For nonprecessing binaries, outgoing gravitational
radiation during the inspiral phase is generated using an ansatz for resumming the post-Newtonian expressions for
outgoing radiation including non-quasicircular corrections, for the leading-order $\ell=2$ subspace.  For the  merger phase of nonprecessing binaries,  the
gravitational radiation is generated via a resummation of many quasinormal modes, with coefficients chosen to ensure smoothness.
The final BH's mass and spin, as well as some parameters in the nonprecessing inspiral model, are generated via calibration to numerical relativity simulations of BBH mergers.
For precessing binaries, building off the post-Newtonian ansatz of separation of timescales and orbit
averaging~\cite{1996PhRvD..54.4813W,1995PhRvD..52..821K,ACST,BCV:PTF}, gravitational radiation during the inspiral is modeled as if from an  instantaneously nonprecessing
binary (with suitable nonprecessing spins), in a frame in which the binary is not precessing~\cite{2011PhRvD..84b4046S,gwastro-mergers-nr-Alignment-ROS-CorotatingWaveforms,2013PhRvD..87j4006B}.
During the merger, the radiation is approximated using the same final BH state, with the same precession
frequency.\footnote{This choice of merger phase behavior is known to be inconsistent with precessional dynamics during merger~\cite{2013PhRvD..87d4038O,2017PhRvD..96b4058B}.}
With well-specified initial data in the time domain, this method can be  directly compared to  the trajectories~\cite{2015PhRvD..92j4028O}
 and radiation~\cite{2017PhRvD..95b4010B}  of numerical BBH spacetimes. 
In this work we use \SEOBA{}, a model for nonprecessing binaries \cite{2017PhRvD..95d4028B}; \SEOBAROM{}, a fast
surrogate model for \SEOBA{} \cite{2017PhRvD..95d4028B}; and
\SEOBP{}, a model for precessing binaries \cite{2014PhRvD..89f1502T,2017PhRvD..95b4010B}.  
The \IMRP model is a part of an approach that attempts to  approximate the leading-order gravitational wave radiation using phenomenological fits
to the Fourier transform of this radiation, computed from numerical relativity simulations and post-newtonian calculation~\cite{nr-Jena-nonspinning-templates2007,gwastro-nr-Phenom-Lucia2010,gwastro-mergers-IMRPhenomP}.  Also using information
about the final BH state, this phenomenological frequency-domain approach matches standard approximations for
the post-Newtonian gravitational wave phase to an approximate, theoretically-motivated spectrum characterizing merger
and ringdown. 
For \IMRP, precession is also incorporated by a ``co-rotating frame'' ansatz, here implemented via a stationary-phase approximation
to the time-domain rotation operations performed for \SEOBP.
Surrogate
models provide  efficient and  accurate representations of  the gravitational wave strain, by interpolating between
evaluations of a costly reference model.  They have been applied to long duration signals~\cite{2014PhRvX...4c1006F,2014CQGra..31s5010P}, arbitrarily many
harmonic modes~\cite{2014PhRvX...4c1006F,2015PhRvL.115l1102B}, spinning binary
systems~\cite{2014CQGra..31s5010P,2017PhRvD..95j4036L},  and neutron star models with tidal
effects~\cite{2017PhRvD..95j4036L}.  In this work, we use surrogates developed to reproduce multimodal radiation from  precessing binary
systems~\cite{2017PhRvD..95j4023B,2017PhRvD..96b4058B,2016PhRvD..94d4031S}.  These  surrogates
  are demonstrably much more
accurate in their domain of validity than the approximations described above.  However, due to the limited duration and
parameter space coverage, the surrogate we use here (\NRSur{},\cite{2017PhRvD..96b4058B}) has a limited range of validity, to high masses; to mass
ratios $m_1/m_2 <2$; and to spins $|\chi_i|<0.8$.

Models for binary neutron stars  account for the response of each neutron star's structure  into their estimate for
dynamics of and radiation from merger.
Using frequency domain methods which generalize classic post-Newtonian calculations
\cite{2008PhRvD..77b1502F,2011PhRvD..83h4051V}, Dietrich et al \cite{2017PhRvD..96l1501D} introduced a simple way to add leading-order tidal effects to the
nonprecessing models described above.  Recently  Dietrich et al \cite{nrtidal}  implemented this effect into 
\SEOBAROM{} and \IMRPD{}, which we denote by  the postfix \textsc{\_NRTidal}. 
These \textsc{\_NRTidal} approximations were developed and calibrated using an effective one body model  (\textsc{\Resum}) that incorporates the effects of
adiabatic tides and spin   \cite{TEOBResumS-2018,2015PhRvL.114p1103B}; see \cite{nrtidal} and references therein.  %
Another model (\textsc{\TEOB}) approximates binary inspiral and neutron star dynamics by allowing  neutron stars to have
both spin and  dynamical tides \cite{2017PhRvD..95d4028B,2016PhRvL.116r1101H,2016PhRvD..94j4028S,SEOBNRv4T-TechnicalNote}.  
As described in part in \cite{nrtidal}, these approximations include tidal effects at differing levels of completeness.
For example, among the models described here only \textsc{\Resum{}} and \textsc{\TEOB{}} incorporate
the  quadrupole-monopole interaction \cite{1998PhRvD..57.5287P} into the dynamics; only \textsc{\Resum{}} includes   higher modes into the outgoing
radiation; and only \textsc{\TEOB{}} uses dynamic tides. 
While none of these models can account for neutron star spins that are misaligned from the orbital angular momentum,
these models capture the leading-order features most relevant for small NS spins.

For most models, we employ  \texttt{lalsimulation} implementations of these two
approximations, provided and maintained by their authors in the same form as used in LIGO's O1 and O2 investigations.
For NR surrogate waveforms, we employ the software provided by the authors \cite{nr-surrogates-web-sxs}.
We use the implementation of \Resum{} provided by the authors \cite{teobresums-code}, restricting to  $\ell=2$ modes for
the outgoing radiation.

Rather than rely on one of these diverse approximations for our synthetic fiducial sources, we instead almost always use the output of
detailed simulations of Einstein's equations, provided by the RIT and SXS group 
 \cite{SXS:catalog,gr-nr-CornellCaltechCatalogDump-2013,2017CQGra..34v4001H}.  
Table  \ref{tab:SourceParameters} provides the properties  of our synthetic sources.
All of the synthetic sources have parameters qualitatively consistent with the observed binary black hole population. 

\begin{table*}
\begin{ruledtabular}
\begin{tabular}{llrrrrrrrrrr}
    ID & Model/Numerical Relativity & $q$ & $M$ ($M_{\odot}$) & $\chi_{\rm 1x}$ & $\chi_{\rm 1y}$ & $\chi_{\rm 1z}$ & $\chi_{\rm 2x}$ & $\chi_{\rm 2y}$ & $\chi_{\rm 2z}$ & $\lambda_1$ & $\lambda_2$\\ \hline
    RIT-1 & RIT:D12\_q1.00\_a-0.25\_-0.25\_n100 & 1.00 & 80.0 & 0.0 & 0.0  & -0.250 & 0.0 & 0.0 & -0.250 & - & -\\
    SXS-1 & SXS:BBH:0308 & 1.23 & 110.0 & 0.117 & 0.0 & 0.320 & 0.335 & 0.0 & -0.580 & - & -\\
    RIT-2 & RIT:De10\_D10.97\_q1.3333\_a-0.6\_-0.8\_n100 & 1.33 & 80.0 & 0.0 & 0.0  & -0.600 & 0.0 & 0.0 & -0.800 & - & -\\
    Tides-1 & \TEOB & 1.11 & 2.88 & 0.0 & 0.0 & 0.0 & 0.0 & 0.0 & 0.0 & 207 & 409 \\
\end{tabular}
\caption{\label{tab:SourceParameters}\textbf{Parameters of synthetic sources}:  This table shows the parameters of all
  the synthetic sources (waveform approximant models and numerical relativity) used in this paper.  $q$ is the mass
  ratio defined with $q>1$ (see Eq. \ref{eq:q}), $M$ is the detector-frame total mass, and $\chi_{*}$ are the components of the
  normalized spins (see Eq. \ref{eq:chii}). All luminosity distances are set such that the signal-to-noise ratio is around 20 (SNR$\sim$20). Other extrinsic parameters are the following: inclination angle from the line-of-sight is $\iota=\pi/4$, right ascension is RA=0.57, declination is DEC=0.1, and the polarization angle is $\psi=\pi/4$
}
\end{ruledtabular}
\end{table*}

\section{Validation examples}
\label{sec:Validation}

In this section, we use both LI and RIFT to infer the parameters of three synthetic sources provided in Table
\ref{tab:SourceParameters}, using the models described in Section \ref{sec:Models}.  
First and foremost, we use these examples to demonstrate both LI and RIFT produce equivalent results when used to
analyze the same synthetic data with the same underlying model.   To compactly characterize both the analysis method and
model, we will use a prefix of LI or RIFT and a postfix of the model name; for example, a LI-\SEOBAROM{} analysis was
performed with LI using the nonprecessing model \SEOBAROM{}.  
We also use these examples as an opportunity to further demonstrate how systematic differences between models can produce
moderately different conclusions about each binary's properties. In that context, we emphasize the contrast with other models and \NRSur{}, which contains higher modes.

Each of the synthetic sources used in this work are created with the ``zero noise realization'', such that the synthetic
detector data is exactly equal to the expected response due to our synthetic source.  For our binary black hole
investigations, we assume gaussian detector noise in a two-detector,
characterized by the same noise strain power spectrum adopted in our analysis of GW150914 in this and prior work
\cite{NRPaper}.   Four our binary neutron star investigation, we assume gaussian noise in a three-detector 
 network (Hanford/Livingston/Virgo) operating at design sensitivity.  
Each source has been scaled to have  an SNR $\sim20$.

\subsection{Aligned NR Source}
\label{subsec:aligned}

\begin{figure*}
\includegraphics[width=\columnwidth]{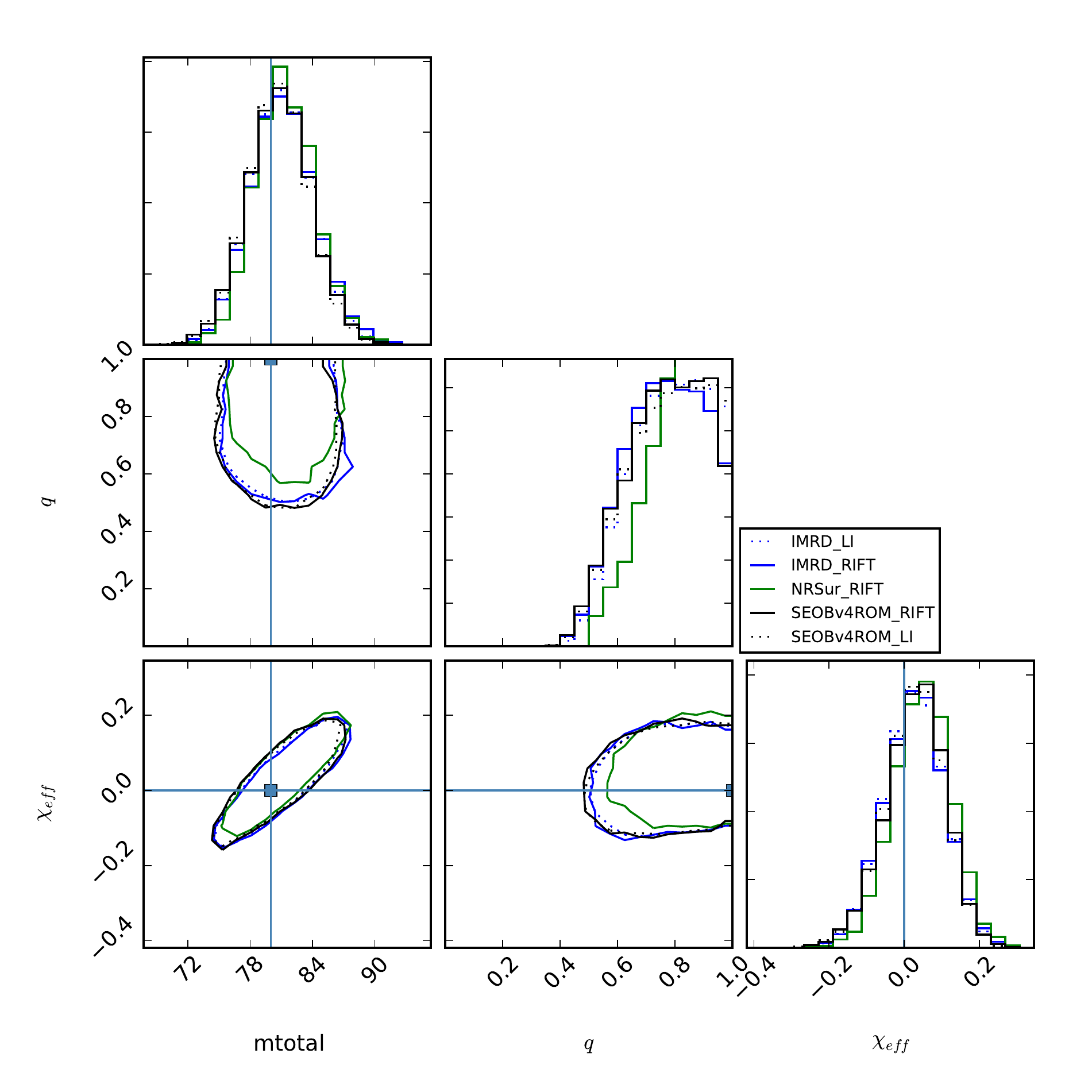}
\includegraphics[width=\columnwidth]{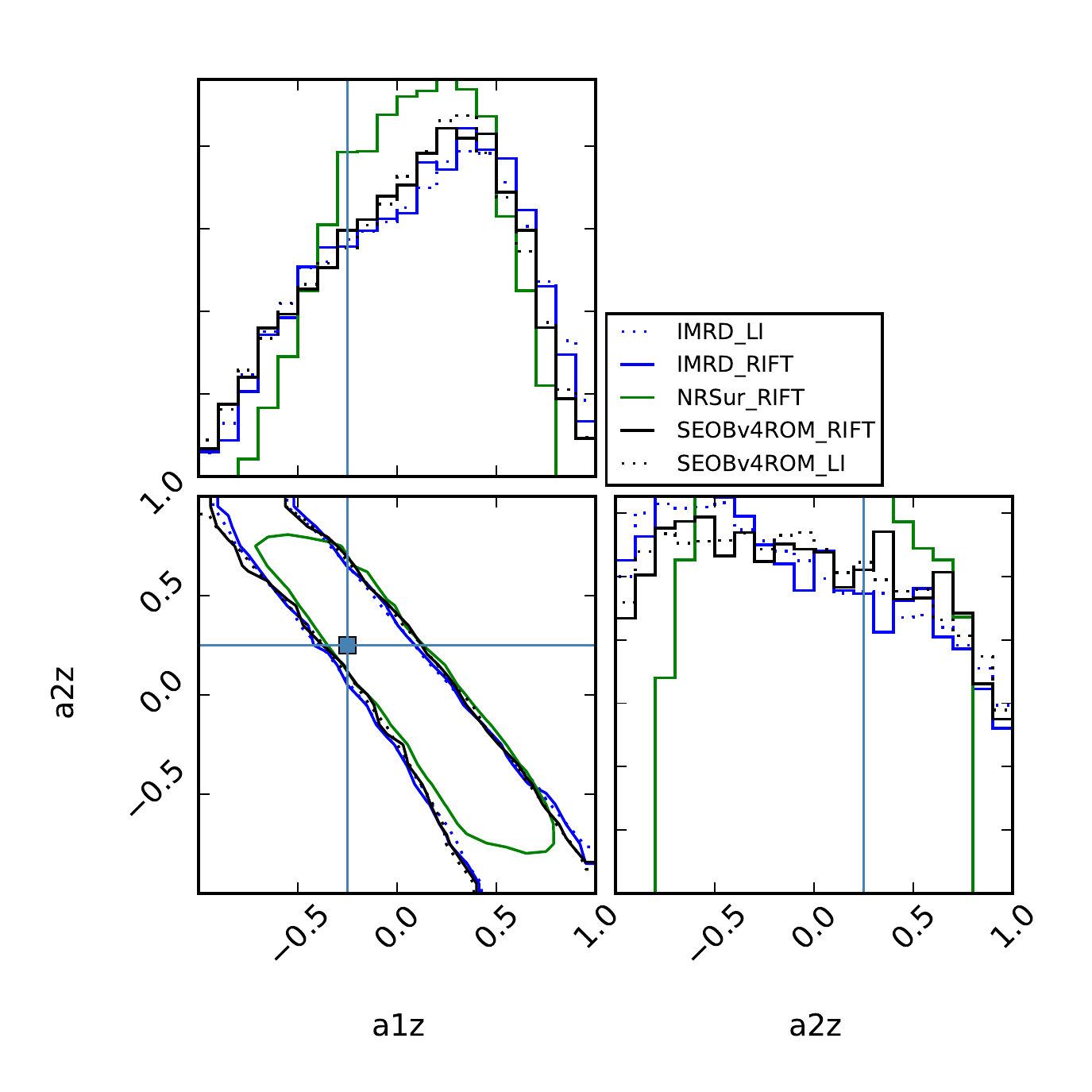}
\includegraphics[width=\columnwidth]{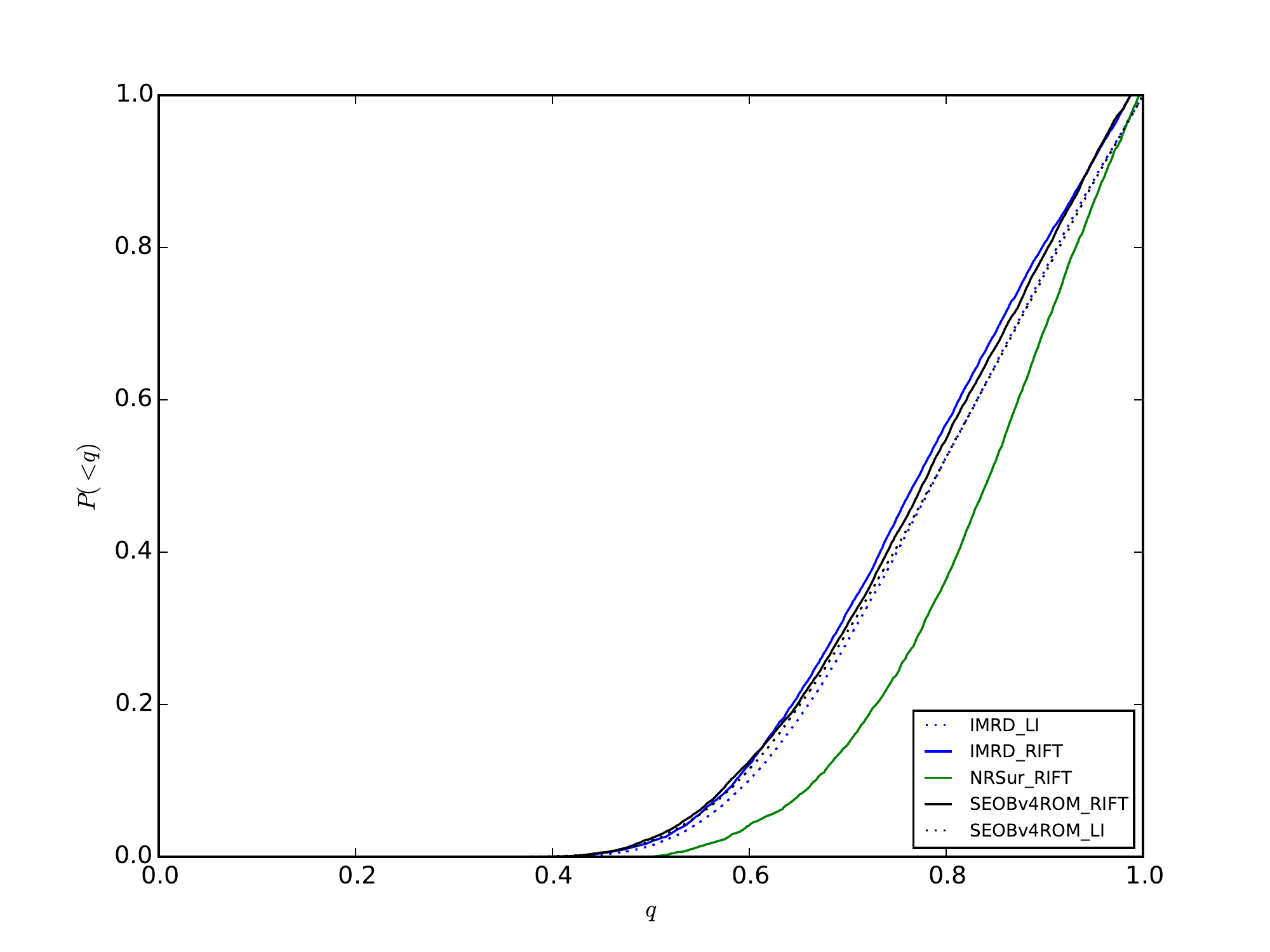}
\includegraphics[width=\columnwidth]{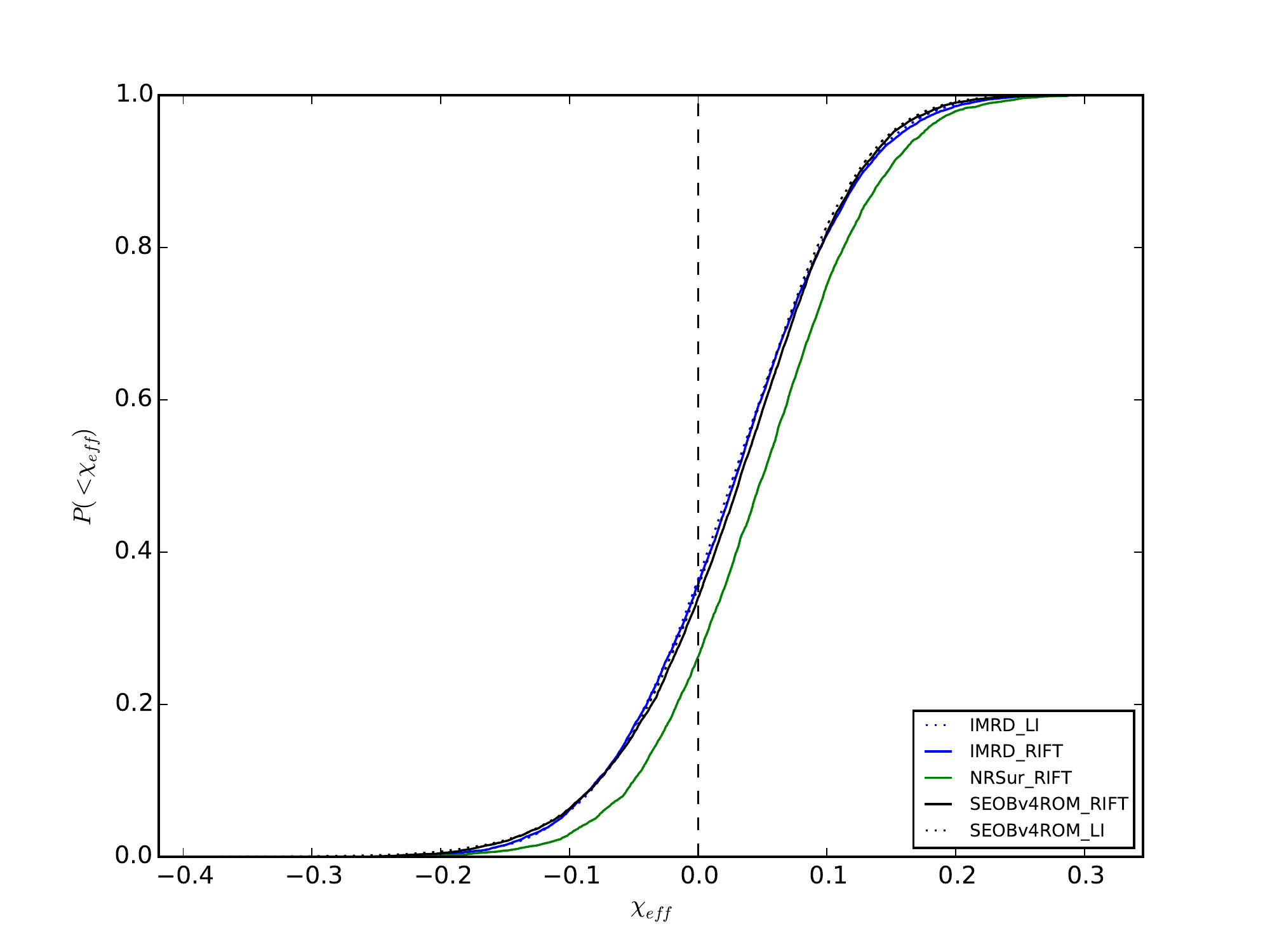}
\caption{\label{fig:aligned1}\textbf{Recovery of nonprecessing NR source with  nonprecessing analyses}. This figure shows
inferred posterior parameter distributions for source \#1 in Table \ref{tab:SourceParameters}, generated using the
\textsc{corner} package \cite{corner}.  The top-left panel shows the
one- and two-dimensional marginal distributions for $M,q,\chi_{\rm eff}$; the top-right panel shows the one- and
two-dimensional marginal distributions for $\chi_{1,z},\chi_{2,z}$. The bottom two panels show the cumulative distributions of parameters $q$ and $\chi_{\rm eff}$.
In the two-dimensional plots, dotted contours are the 90\%
  confidence intervals for the LI results using the \IMRPD{} and \SEOBAROM{} models, while the three solid curves show \RIFT{} results
  produced with \IMRPD (blue, solid), \SEOBAROM{} (black, solid) and \NRSur{}, restricted to aligned spins and including all modes up to $\ell=3$
  (green, solid). %
Both LI and \RIFT{} produce comparable results when using \IMRPD{} or \SEOBAROM, with statistical differences far smaller than model
systematic effects (illustrated here with \NRSur{}). 
}
\end{figure*}

The first synthetic source corresponds to the first row of Table \ref{tab:SourceParameters} (RIT-1): an equal-mass
binary black hole with $\chi_{1,z}=\chi_{2,z}=-1/4$ and $M=80 M_\odot$.  
For this analysis, we assume both  spins are parallel to the orbital angular momentum, and  adopt a uniform prior
distribution in $\chi_{i,z}$ from the maximum to minimum allowed by each approximation. 
Using \RIFT{}, we estimate the parameters with the following nonprecessing models: \SEOBAROM{}, \IMRPD{}, and \NRSur{} with the latter using modes with $\ell \le 3$.  Using LI, we estimate the
parameters with the same two former models \SEOBAROM{} and \IMRPD{}. 
Figure \ref{fig:aligned1} shows our inferred 90\% credible intervals for several parameters for each of these algorithms
and models, as well as associated one-dimensional
marginal posterior distributions.

First and foremost, Figure \ref{fig:aligned1} demonstrates that LI and \RIFT{} produce comparable results when analyzing
this data with 
\SEOBAROM{} (LI is dotted black, \RIFT{} is solid black) as well as with \IMRPD{} (LI is dotted blue, \RIFT{} is solid blue).  %
Second,  due to  the well-known good agreement between the nonprecessing models \SEOBAROM{} and \IMRPD{} for the (2,2)
mode \cite{gwastro-mergers-nr-PrayushAlignedSpinAccuracy-2016,2017PhRvD..95d4028B},  we find very consistent posterior
distributions for \SEOBAROM{} (black) and \IMRPD{} (blue).
Finally and also as expected, because these two models omit higher-order modes, an analysis with an NR surrogate model which includes
several higher-order modes (here, $\ell_{max}=3$) draws sharper conclusions about at least one of the parameters: the
binary mass ratio.  [Conclusions about the other parameters are not significantly impacted by including higher-order modes.]
A similar result was found with a reanalysis of GW150914 where the LVC compared the data directly to NR \cite{NRPaper}.

\subsection{Precessing NR Source}
\label{subsec:precess}
\begin{figure*}
\includegraphics[width=\columnwidth]{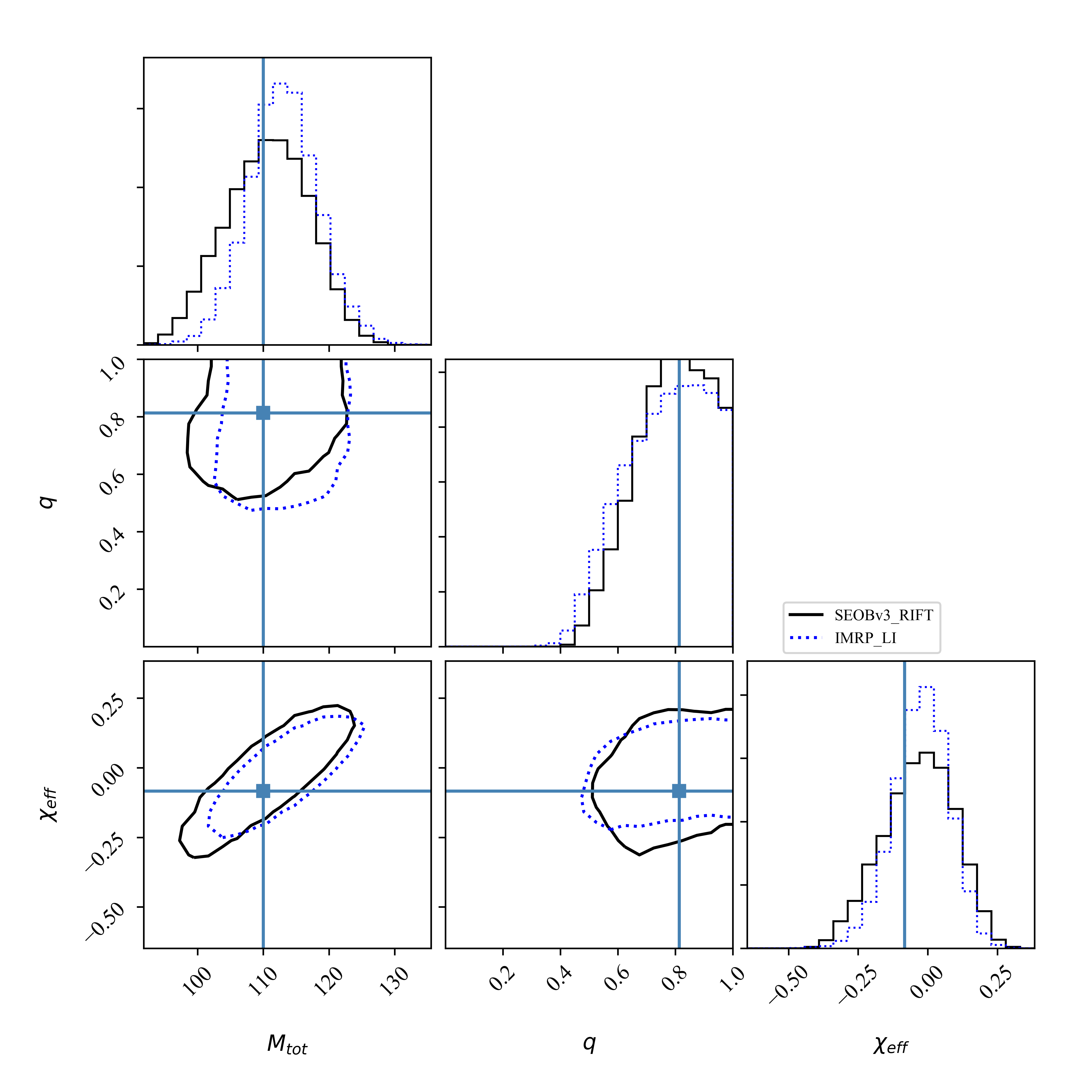}
\includegraphics[width=\columnwidth]{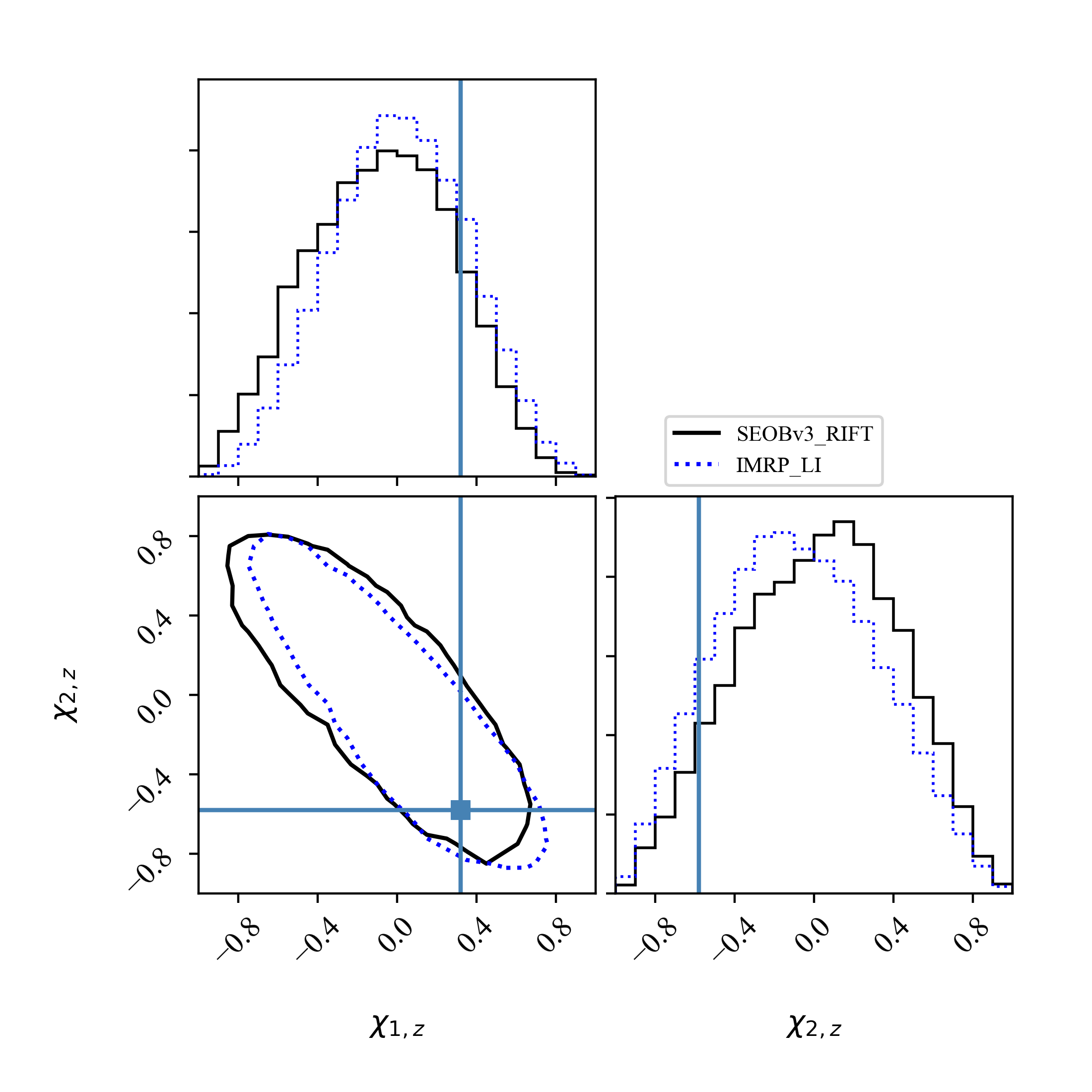}
\caption{\label{fig:precess1}\textbf{Recovery of a precessing NR source with a precessing model}. This figure shows the
  results using source \#2 in Table \ref{tab:SourceParameters}. The first group of  plots shows the PE results in the
  total mass ($M_{\rm tot}$)-effective spin
  ($\chi_{\rm eff}$)-mass ratio (q) parameter space (see Eq. \ref{eq:chieff} and Eq. \ref{eq:q} respectively). The second
  group of plots shows inferences about the two spin components $\chi_{i,z}$, where the $z$ axis convention is defined
  as parallel to the orbital angular momentum.  
}
\end{figure*}

To test our approach on a precessing source, we generated a synthetic precessing signal using the second row of Table
\ref{tab:SourceParameters} (SXS-1), corresponding to a comparable-mass system with modest spins perpendicular to the orbital
plane.   %
We  adopt a volumetric prior on each BH's spin; see Appendix \ref{subap:spins}. 
Using RIFT, we again estimate the parameters but this time with the fully
precessing time-domain \SEOBP{} and \NRSur{} models. Using LI, we infer source properties
 with the precessing frequency-domain \IMRP{} model. %
Figures \ref{fig:precess1} show our inferred 90\% confidence interval for each model,  again with 
LI represented with dotted lines; RIFT results with solid lines; \SEOBP{} in black; and IMRP in blue.

As expected, due to well-known differences between \IMRP{}  and  \SEOBP{}, for example as demonstrated in
\cite{2017PhRvD..96l4041W}, the conclusions derived using these two models do not perfectly agree. 

\subsection{Binary neutron star}
\label{sec:sub:BNS}

\begin{figure*}
\includegraphics[width=\columnwidth]{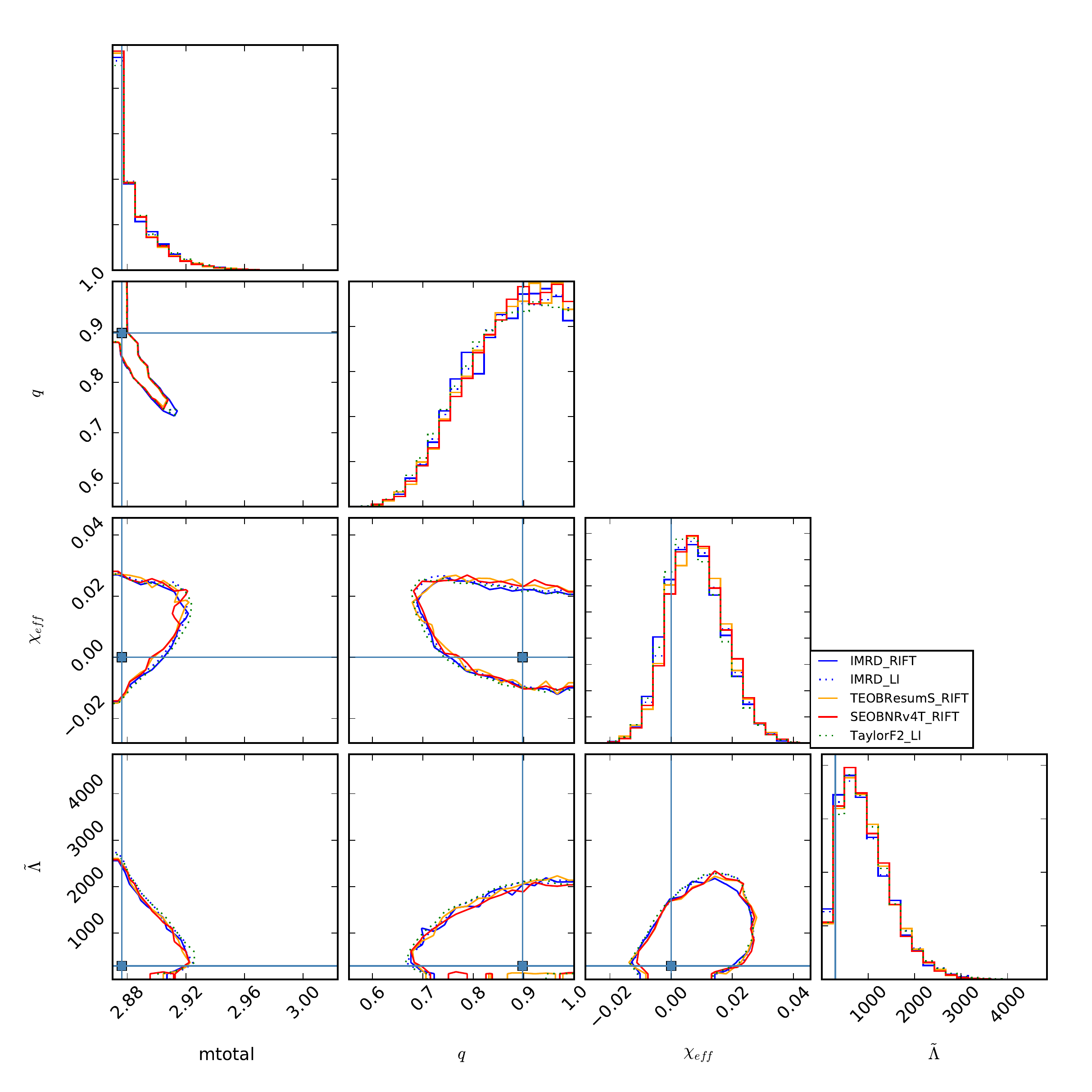}
\includegraphics[width=\columnwidth]{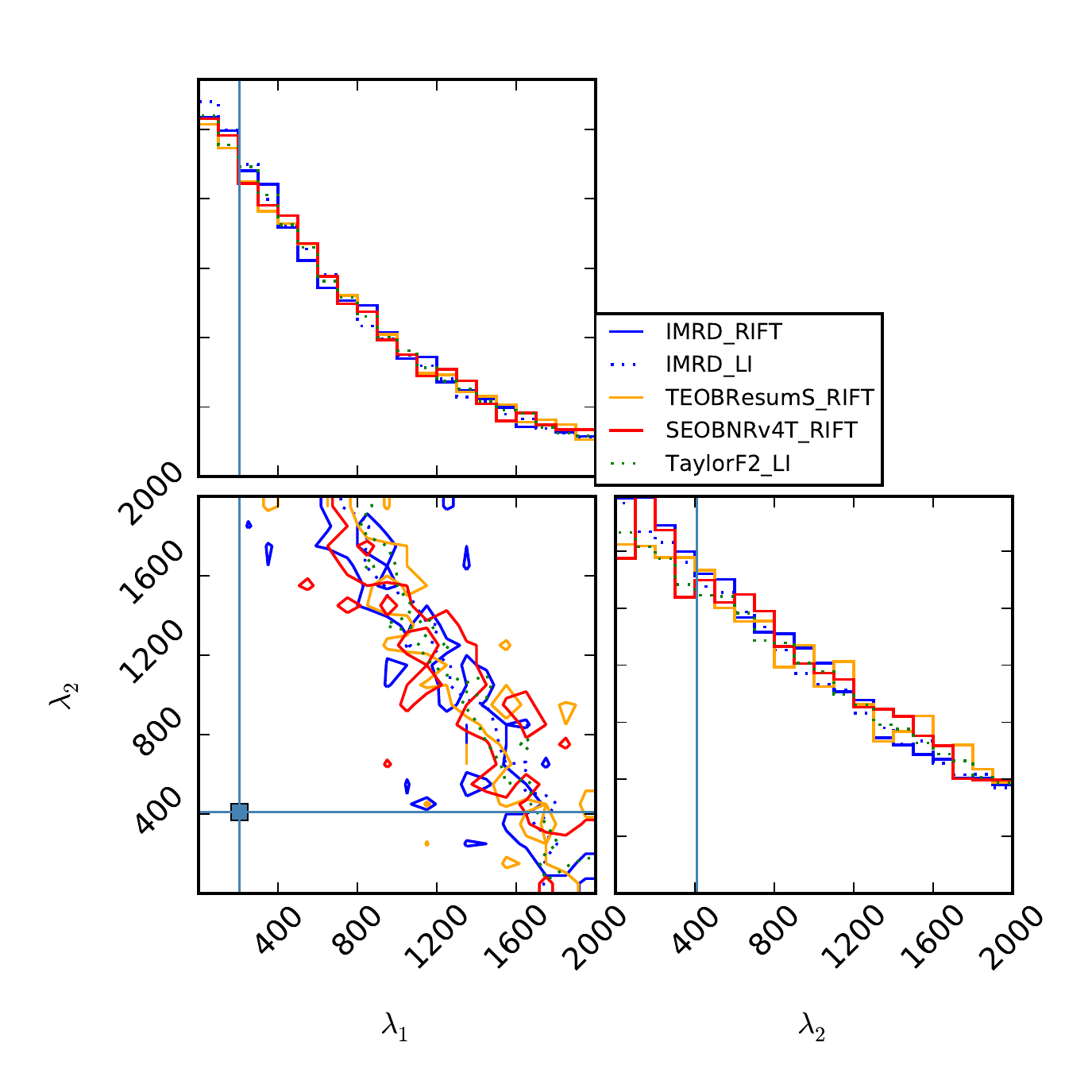}
\includegraphics[width=\columnwidth]{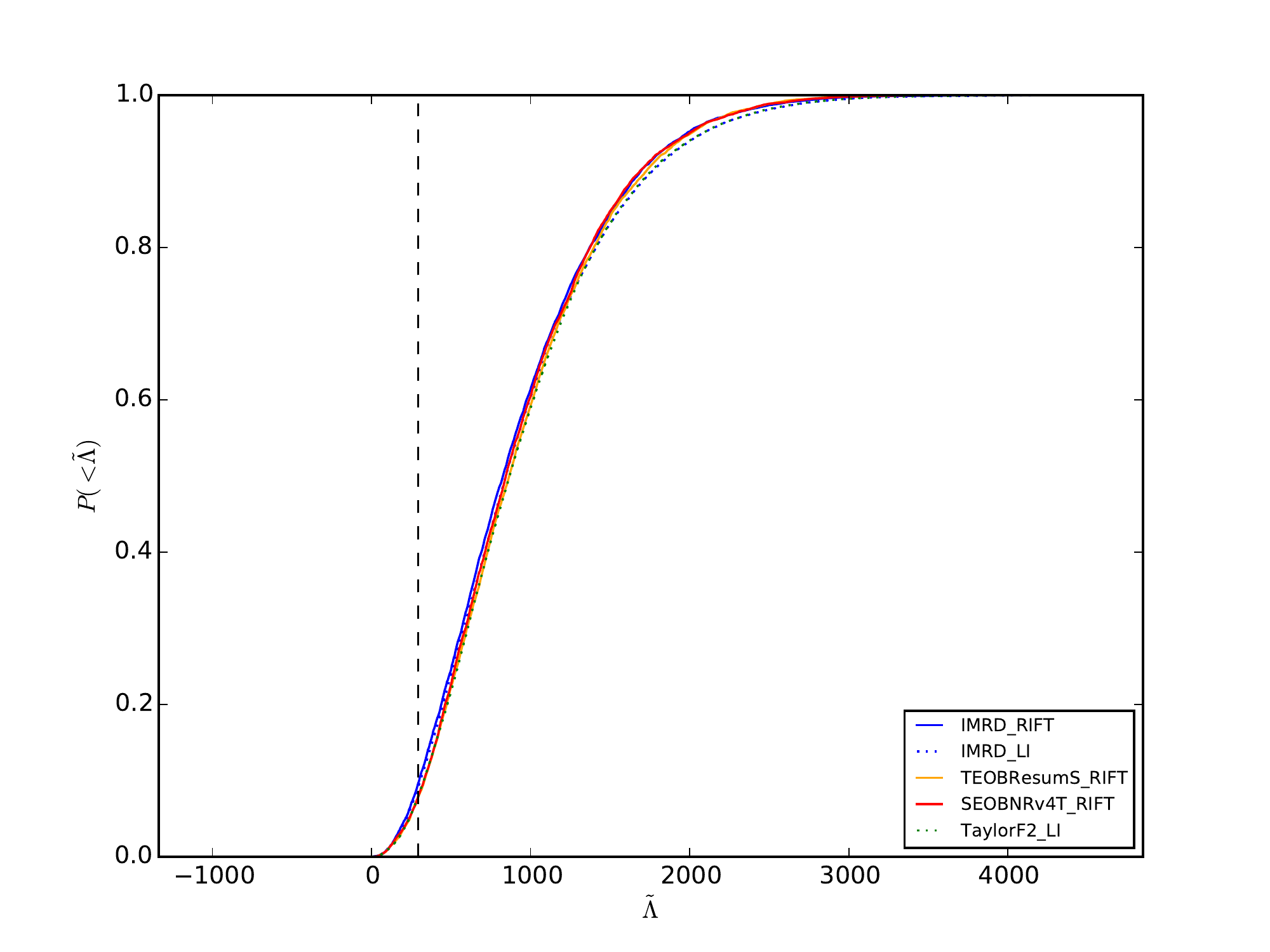}
\includegraphics[width=\columnwidth]{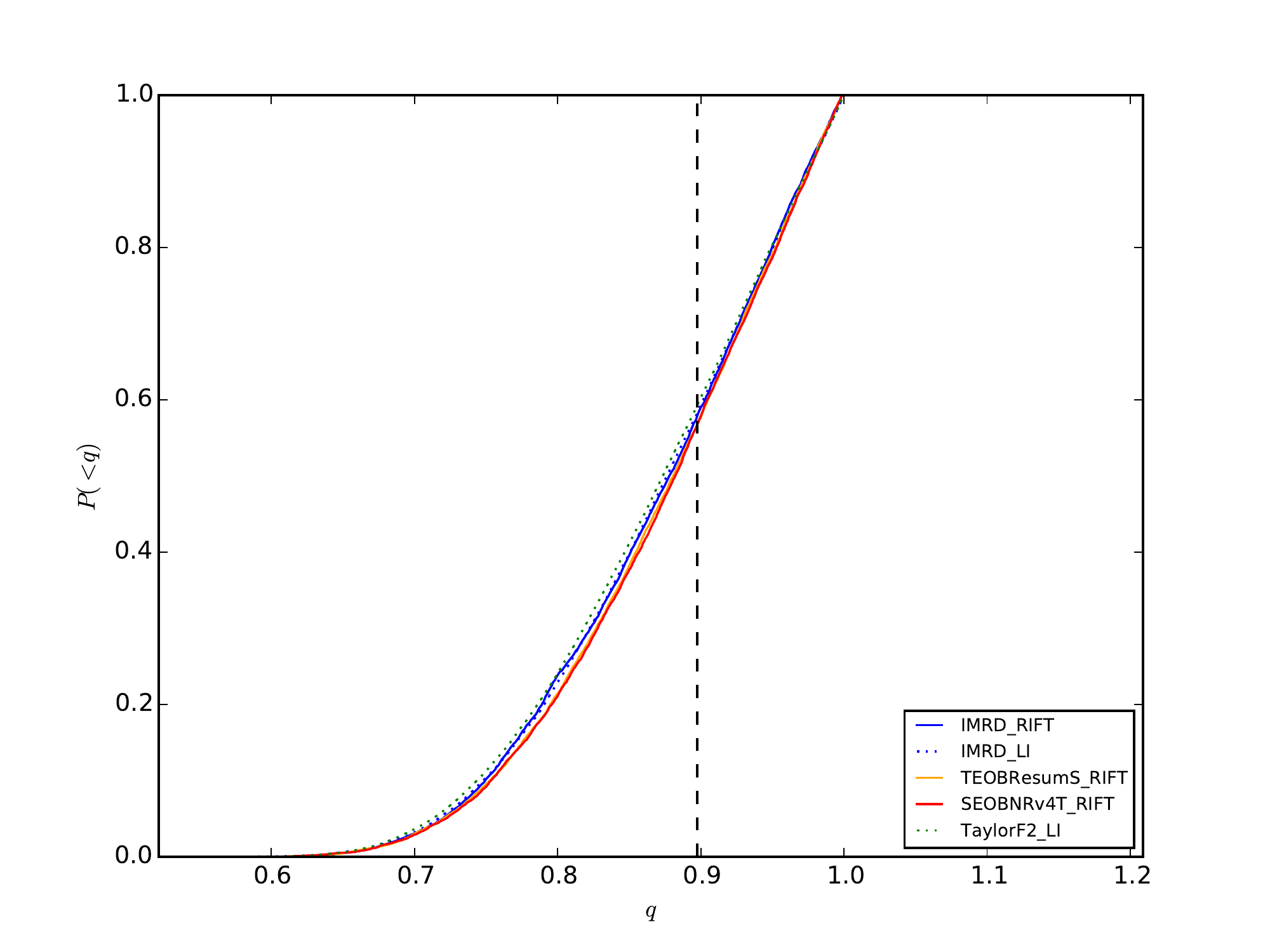}
\caption{\label{fig:tides1}\textbf{Using sophisticated tidal waveforms to analyze tidal injections}  This figure
  shows inferences about the Tidal-1 source Table \ref{tab:SourceParameters}, performed
  adopting the spin prior in Eq. (\ref{eq:zprior}) with $\chi_{\rm max}=0.05$ . As in
  previous figures, the one- and two-dimensional results show all possible one-dimensional posterior distributions and
  two-dimensional 90\% credible intervals for the total mass ($M$), mass ratio ($q$), effective tidal deformability
  ($\LambdaTilde$), and net effective spin ($\chi_{\rm eff}$).  
As previously, dotted contours and curves correspond to LI results; solid curves are produced with \RIFT{}; and
different colors correspond to different waveform models: TaylorF2 is green, \IMRPDT{} is blue, \TEOB{} is red, and \Resum{} is orange.  
}
\end{figure*}

Due to their low mass, binary neutron stars produce exceptionally long inspirals in the sensitive band of LIGO and
Virgo.  \RIFT{} can not only efficiently analyze these signals, but do so while using exceptionally costly
source models like \TEOB{} and \Resum{}, which can require up to an hour to generate per source.   
As a concrete example, Figure \ref{fig:tides1} presents an analysis of synthetic data based on the Tidal-1 entry in
Table \ref{tab:SourceParameters}.  %
Firstly, the dotted and solid blue contours and distributions show 90\% credible intervals and
one- and two-dimensional marginal distributions inferred from with LI and \RIFT{}, both using the \IMRPD{} waveform model,
modified to include tides \cite{nrtidal}.  Both are consistent.  Secondly, \RIFT{} analyzed the data with the computationally expensive waveform models: \TEOB{} in red and \Resum{} in orange. These again agree nicely with the previously mentioned \IMRPDT{} results as well as the LI-TaylorF2 in green. Using \RIFT{}, we can not only recover the correct posterior, but we can do so with more costly waveforms when analyzing a binary neutron star system.

\section{Applications}
\label{sec:Applications}

\subsection{Parameter inference via mixed models}

Many approximations are well-suited only to certain parts of the parameter space, breaking down for sufficiently extreme
spins, mass ratio, or total mass. For example, \NRSur{} is only suitable for mass ratios $q > 0.5$ and
$|\mathbf{\chi}_i| < 0.8$.   
\RIFT{} provides an almost-trivial mechanism to flexibly explore parameter inferences that employ different
approximations $A,B$ in different regions. Because the computational cost is dominated by ILE, once parameter
inference is performed using models $A$ and $B$ separately, we can reanalyze the data with different mixtures of $A$ and $B$
with no additional overhead. 
For simplicity and to illustrate the method, we will employ the most extreme and simple form of this approach, where the training pairs
$(\mathbf{\lambda}_k,\lnLmarg(\mathbf{\lambda})_k)$ used to produce a fit for $\lnLmarg(\mathbf{\lambda})$ are provided
using model $A$ when $\mathbf{\lambda}_k$ is in some region ${\cal V}$, but with model
$B$ everywhere outside ${\cal V}$. 

As a concrete and practical motivation for this strategy, Figures \ref{fig:aligned1} and \ref{fig:systematics:Motivation}
show analysis of synthetic NR sources
 with \IMRPD{} and with \NRSur{}, assuming a nonprecessing binary and adopting a uniform-$\chi_{i,z}$ prior on the two BH spins.
The hard limits on  \NRSur{} both in mass ratio and particularly in $\chi_{i,z}$ have a substantial impact on our
conclusions about spins and, for systems with large $|\chi_{\rm eff}|$, even about $\chi_{\rm eff}$ and hence about other
parameters of the binary.  Because  constraints on $\chi_{\rm eff}$ strongly correlate with information extracted about
 binary masses, this limitation is not academic, and can complicate attempts to apply these kinds of limited-support
 waveforms to probe effects like waveform systematics. 

\begin{figure*}
\includegraphics[width=\columnwidth]{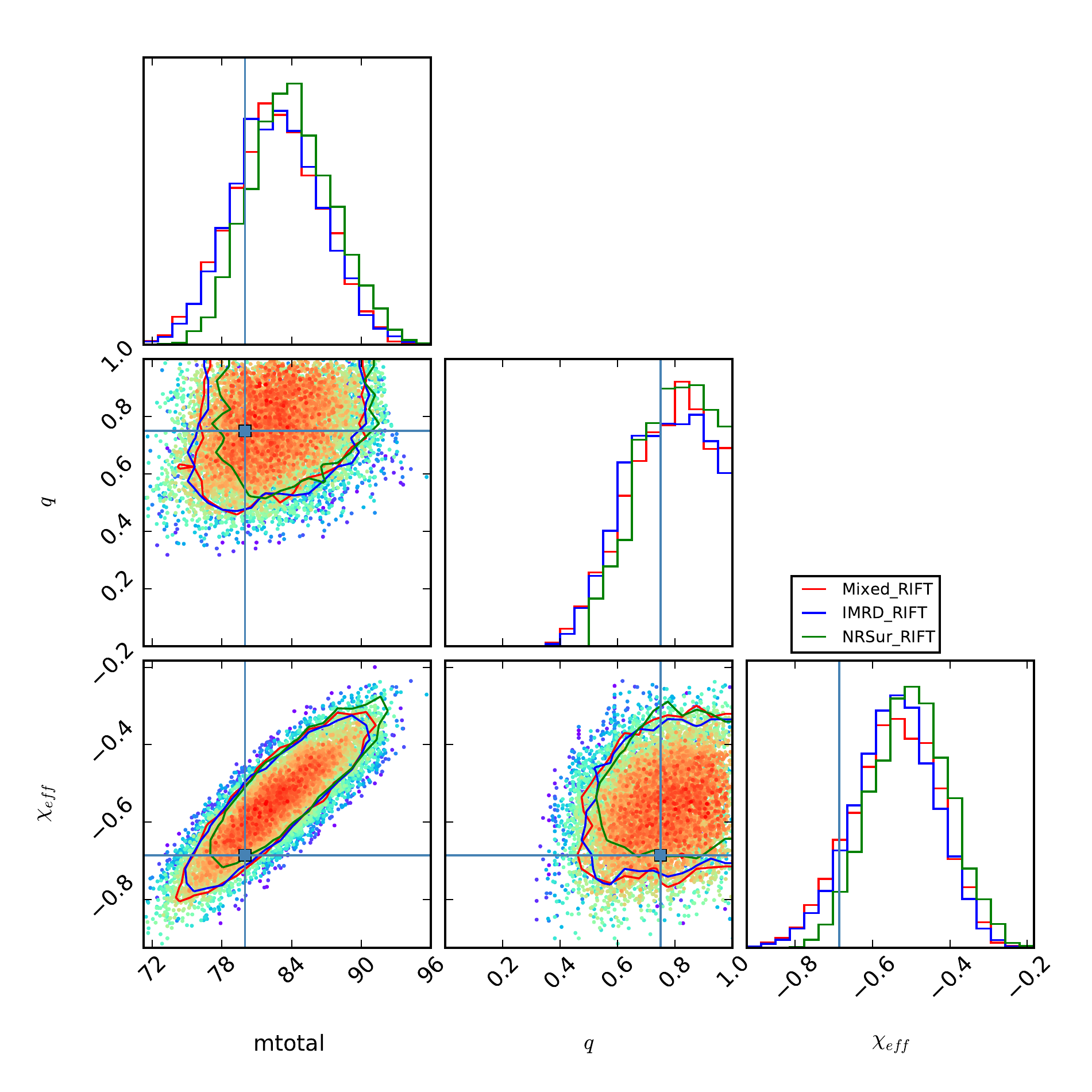}
\includegraphics[width=\columnwidth]{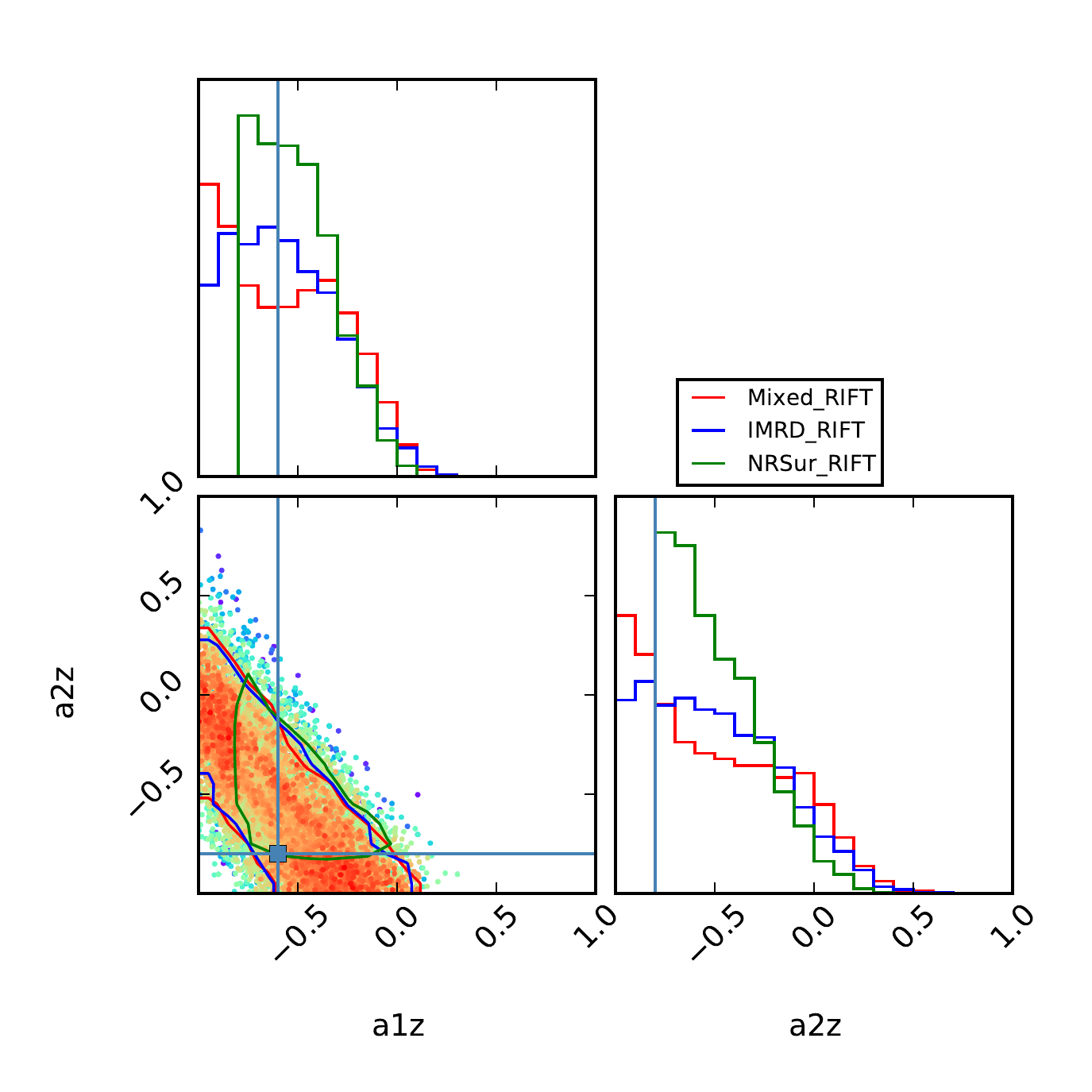}
\includegraphics[width=\columnwidth]{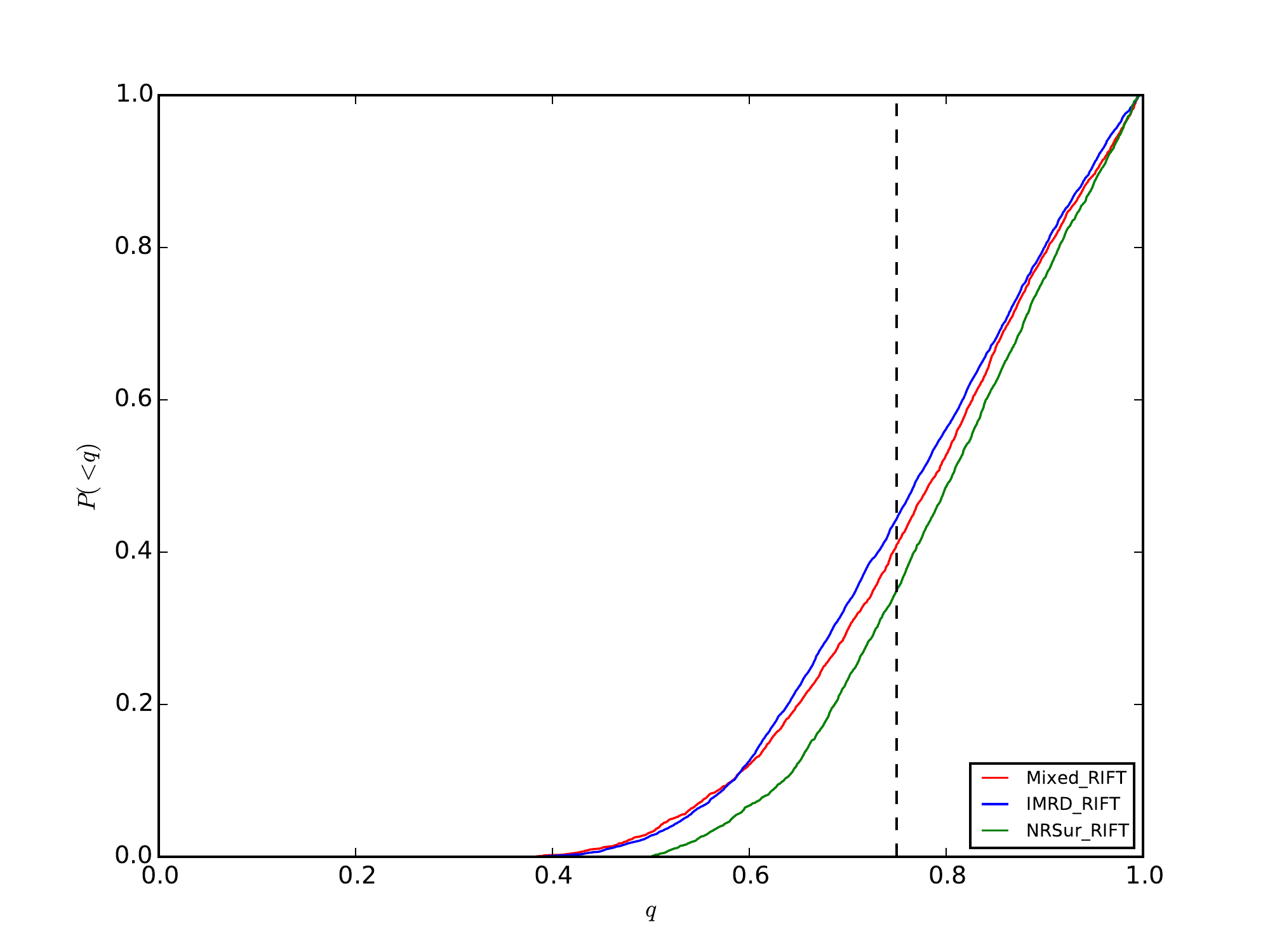}
\includegraphics[width=\columnwidth]{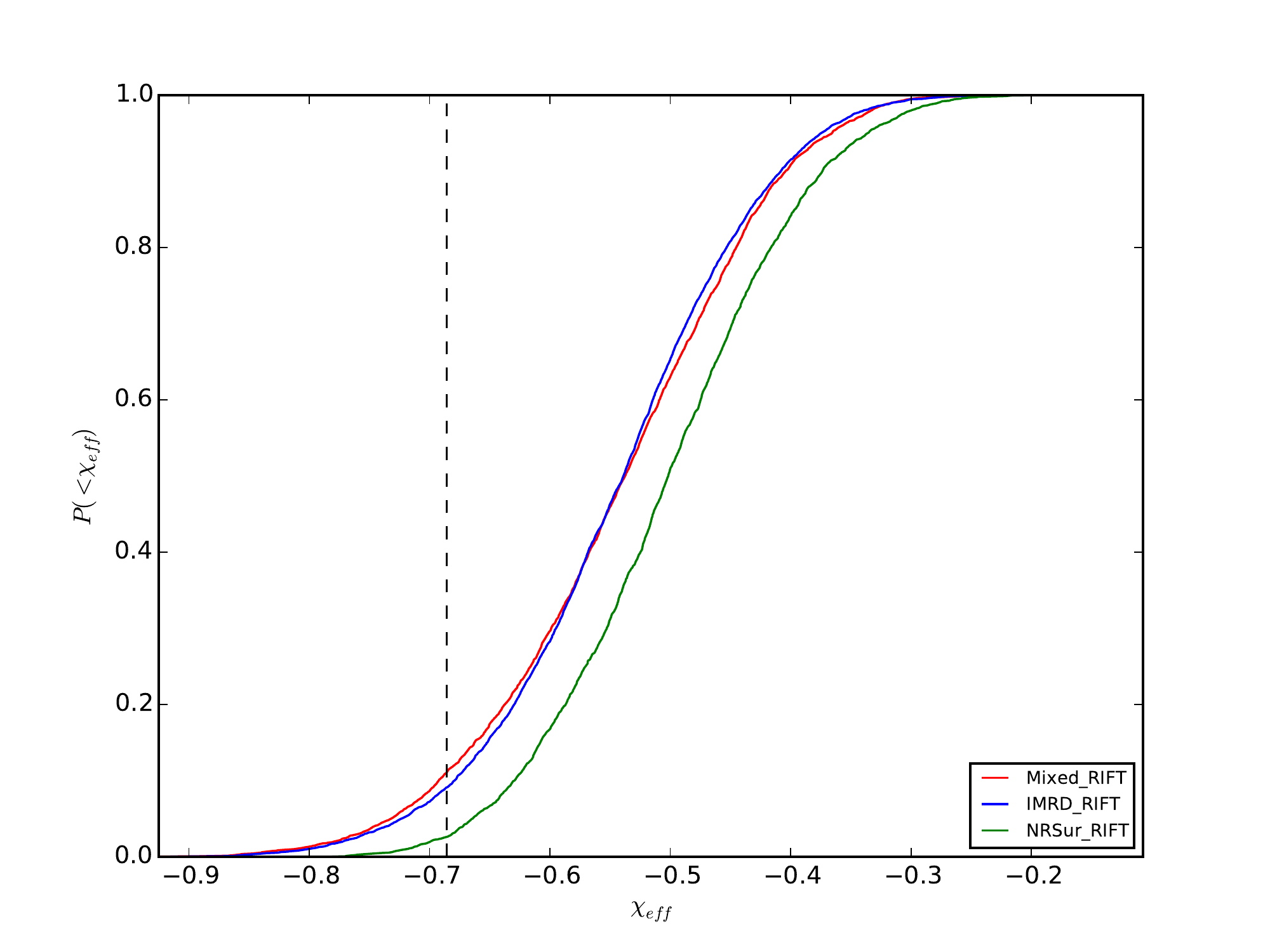}
\caption{\label{fig:mixed1}\textbf{Mixing models for a single analysis}  This figure shows the inferred parameters for
  source RIT-2, in a format similar to Figure \ref{fig:aligned1}.  The colored curves represent 90\% credible
  intervals derived only using \IMRPD{} (blue), \NRSur{} with $\ell=2$ modes only (green), or using a mixture of models (red).  In the mixture approach, we employ
  \NRSur{} with modes up to $\ell_{\rm max}=2$ in its region of validity ($1/q>0.5$ and $a_{i}<0.8$) and \IMRPD{}
  elsewhere.  
}
\end{figure*}

Figure  \ref{fig:mixed1} shows an analysis of the same source (RIT-2) as Figure \ref{fig:systematics:Motivation}, using  \RIFT{} with a hybrid analysis that
uses \NRSur{} with $\ell\le 2$ modes in its domain of validity, and \IMRPD{} elsewhere.   As demonstrated by Figure
\ref{fig:systematics:Motivation}, restricting parameter inference to the  domain of validity of \NRSur{} has a
significant impact on the inferred source parameters.   By contrast, the mixed model result agrees with inferences
adopted with \IMRPD{}.  For the purposes of illustration we have intentionally selected a source both where the domain of
validity of \NRSur{} impacts multiple astrophysical inferences and where extending the model produces comparable
results.  However, as also illustrated with Figure \ref{fig:systematics:Motivation}, even in this extremely
astrophysically pertinent example, these mixed
models are being applied in a regime where \IMRPD{} and \SEOBAROM{} disagree. 
In a subsequent investigation, we will follow up this proof- of-concept example with a more detailed analysis of the
advantages and applications of this mixed-model strategy, particularly for assessing waveform systematics. 

\subsection{Investigating systematics}
The method provides a particularly straightforward but deep method to  assess the impact of  systematics on
posterior distributions.  Simply put, if $A$ and $B$ are two models, we can use precisely the same evaluation points
$\bm{\lambda}_\alpha$ to estimate the marginalized likelihoods according to both models ($\lnL_\alpha(A)$ and
$\lnLmarg_\alpha(B)$), and hence to estimate the corresponding posterior distributions.  In practice, we first generate
a posterior distribution using the iterative procedure described above for model $A$ (e.g., SEOBNRv4) and then, using
all the test points proposed during the iterative scheme, derive the corresponding posterior for model $B$.  This
approach provides not only the posterior distributions but also the ingredients needed for a  detailed investigation
into the origin of any discrepancies: the point-by-point differences between $\lnLmarg_\alpha(A)$ and
$\lnLmarg_\alpha(B)$, as a function of model parameters.   
Once differences are identified, the ability to quickly produce a single scalar diagnostic for model differences
($\lnLmarg$) enables detailed and easily-understood diagnostics as users change one feature of their calculation at a
time (e.g., mode content; data conditioning or noise model).

\begin{figure*}
\includegraphics[width=\columnwidth]{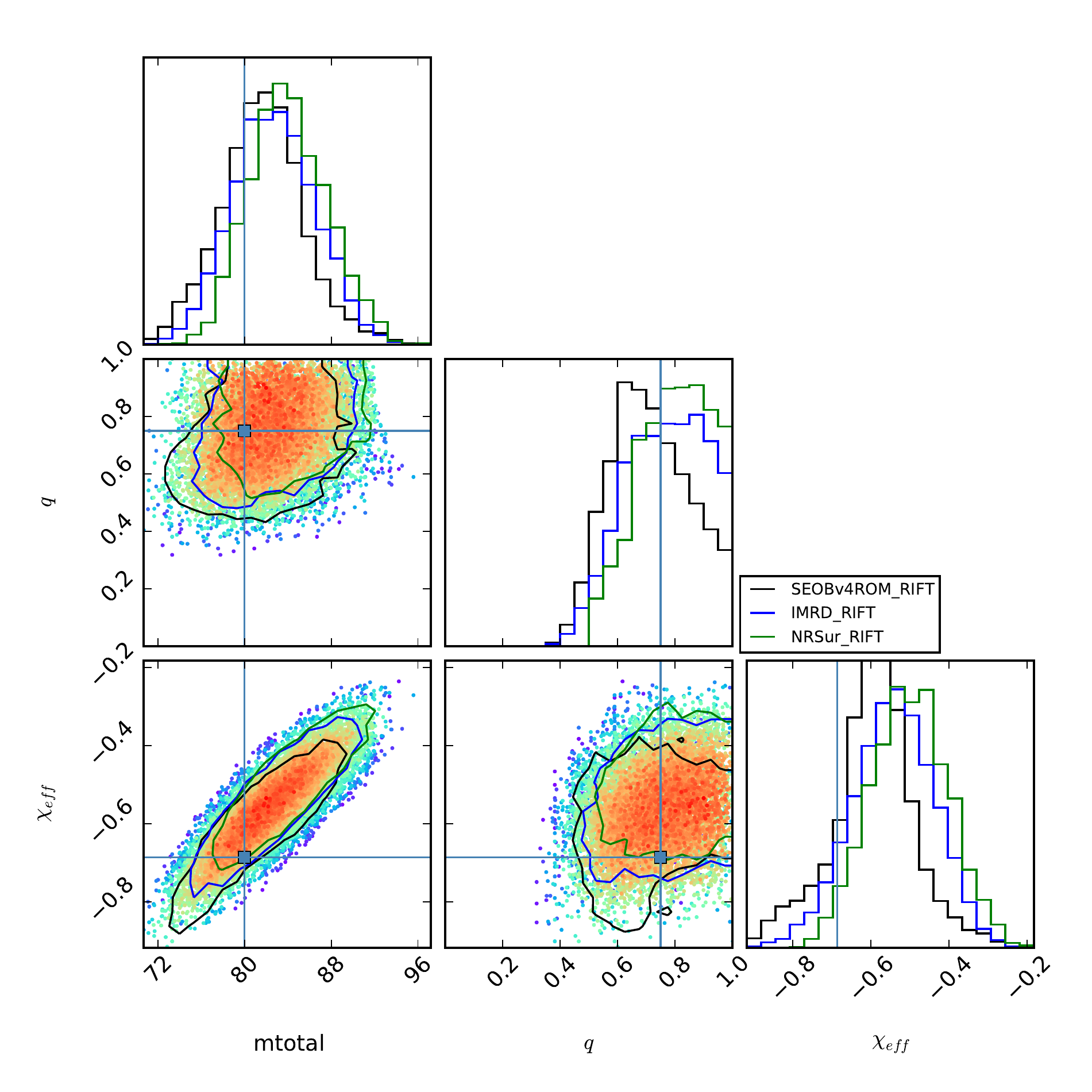}
\includegraphics[width=\columnwidth]{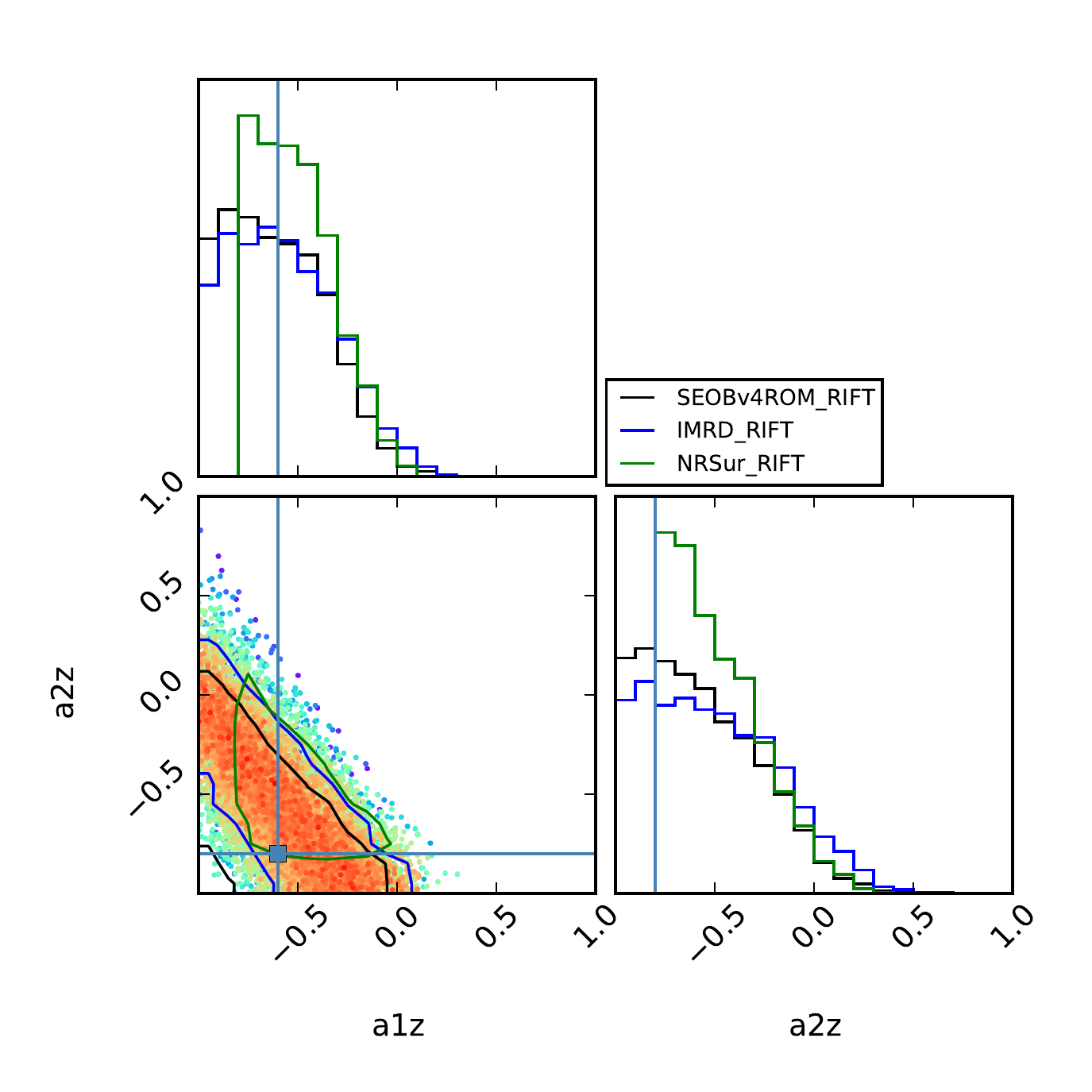}
\includegraphics[width=\columnwidth]{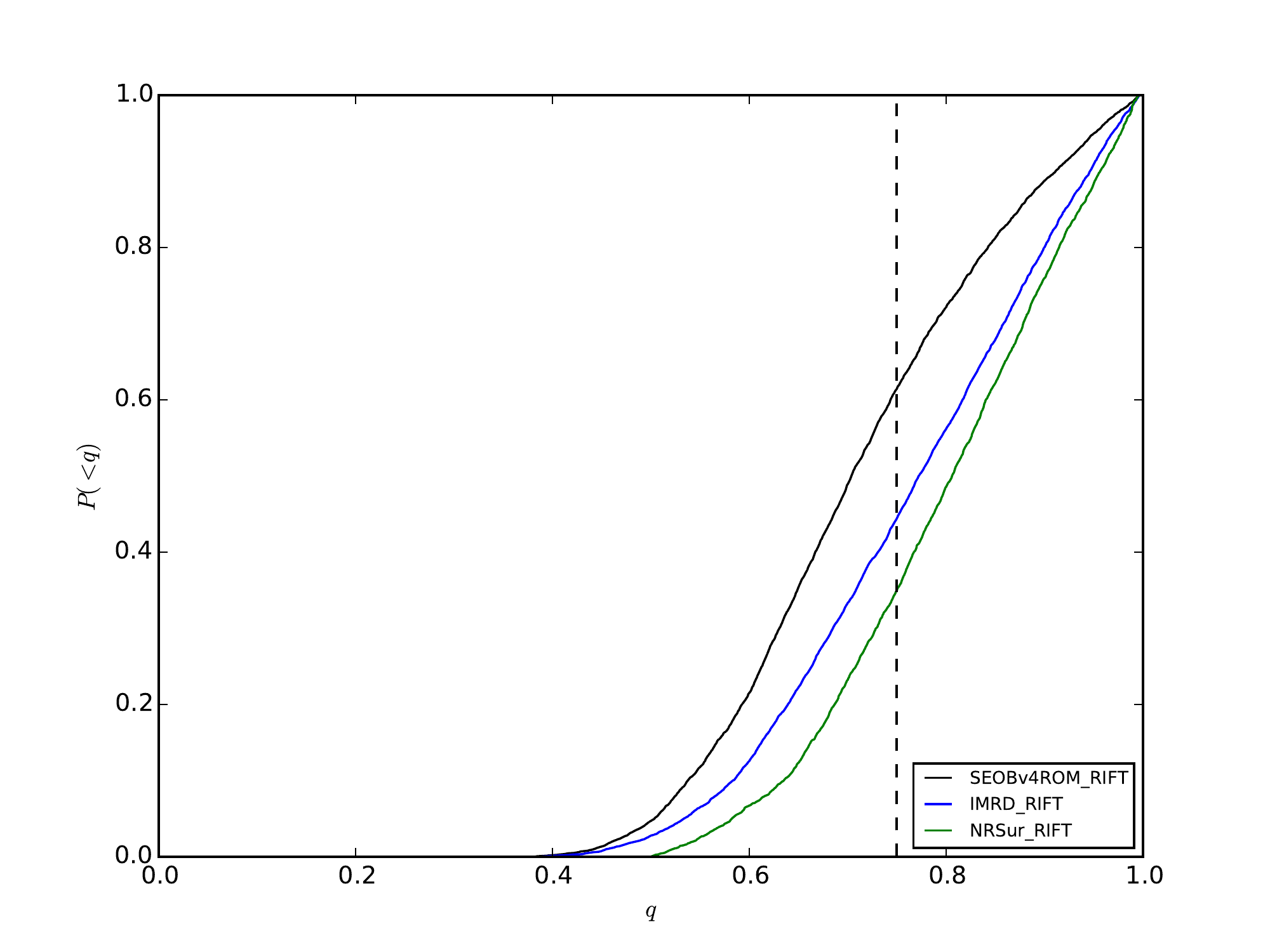}
\includegraphics[width=\columnwidth]{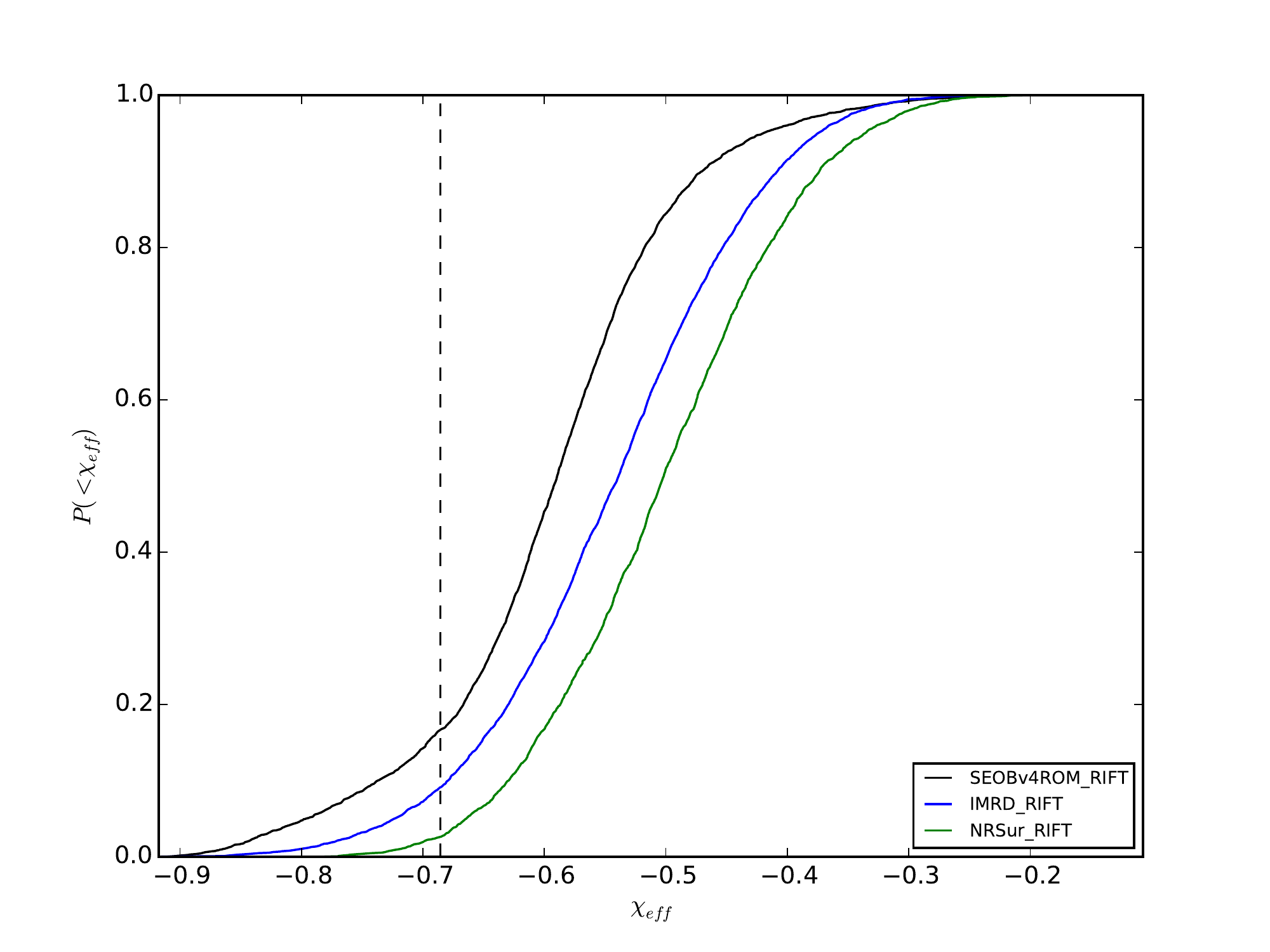}
\caption{\label{fig:systematics:Motivation} 
\textbf{Different models on the same parameter grid} This figure shows inferences about RIT-2, described  in Table
\ref{tab:SourceParameters}. 
The contours, points, and curves used in this figure are styled according to scheme described in Figure
\ref{fig:aligned1}.  Points are colored using likelihood values obtained from \SEOBA.
}
\end{figure*}

Figure \ref{fig:systematics:Motivation} shows a concrete illustration of this strategy, applied when interpreting the
nonprecessing binary source model RIT-2.  In this case, parameter
inference was performed  using a fiducial model to generate a sequence of ever-finer evaluation grids.  The net
grid was then applied to two other models, leading to different predictions.  Because these models all rely on the same
input grid $\bm{\lambda}_\alpha$, we can directly diagnose which features drive differences in our posterior distributions. 
For example, in this case (Figure \ref{fig:systematics:Motivation}) this strategy helps us assess the relative role of
restricted $\chi_i$ versus model differences such as higher modes in changing inferences about $\chi_{\rm eff}$ and hence $M,q$.

\subsection{Astrophysical population inference}
Since \RIFT{} separates the process of estimating the likelihood and producing fair draws from the posterior, it is
ideally suited for astrophysical inference, where all events are reassessed concurrently with different astrophysical
priors.  In this application,   the ability to robustly explore alternative priors is paramount, particularly when future observations may motivate detailed reanalysis of earlier events. 
As described in Section \ref{sec:Methods} and Appendix \ref{ap:CoordinatesAndPriors}, \RIFT{} can
efficiently generate posterior distributions using any prior whatsoever, by performing a suitable Monte Carlo.

In principle, other methods can also reanalyze existing results using alternative priors.  Notably, with a sufficiently large list of posterior samples performed using a prior without
compact support, the procedures  described in  Section \ref{sec:Methods} could produce  weighted samples suitable to any
prior.  However, as emphatically demonstrated by all of our figures (Figs \ref{fig:precess1}) and by Williamson et al
\cite{2017PhRvD..96l4041W}, a finite list of samples from our posterior distribution with fiducial priors have compact support.  For precessing binaries, previous analyses do not sample
extremal spin.  By contrast,  by preserving the full (marginalized) likelihood, \RIFT{} can evaluate the implications of
a nearly arbitrarily extreme choice of prior, with minimal additional computational cost.  
In related work \cite{gwastro-Wysocki-PopModels}, we employ this  approach with \RIFT{} inputs for astrophysical inference.

\begin{widetext}
\subsection{Inferences about the nuclear equation of state}
\label{sec:sub:EOS}
\RIFT{} provides direct estimates of the (marginalized) likelihood $\lnLmarg(\boldsymbol{\lambda})$, allowing us to carry out subsequent calculations
which require it.  As a concrete example, we can more efficiently deduce the nuclear equation of state by post-processing
generic calculations performed by \RIFT{} without added computational expense.    

Specifically, in Section \ref{sec:sub:BNS} and Figure \ref{fig:tides1} we carried out
model-independent inference of binary neutron star masses, spins, and tidal deformabilities $\lambda$.  Any proposed 
nuclear equation of state imposes 
a specific relationship between the (source-frame) neutron star gravitational masses $m_i$ and 
the tidal $\lambda$ parameter ($\lambda(m_i)=\lambda_i$).  Imposing this relationship, we can repeat our parameter
inference both to  draw tighter inferences about binary parameters and to deduce how well the proposed equation of state
matches the data, after marginalizing over all other quantities.  With \RIFT{}, this reanalysis requires a particularly
efficient re-use of the marginalized likelihood $\lnLmarg(m_{1,z},m_{2,z},\lambda_1,\lambda_2,\chi_{1},\chi_2)$ as a
function of its parameters: the redshifted detector-frame masses $m_{i}(1+z)=m_{i,z}$; the two neutron star tidal
deformabilities $\lambda_i$; and the two neutron star spins $\chi_{i}\in[-\chi_{max},\chi_{max}]$, assumed parallel to the orbital
angular momentum.  
Specifically, for each proposed equation of state, characterized by some hyper-parameters $\gamma$, we evaluate the
a quantity equivalent to the evidence
$ \int d\bm{\lambda} {\cal L}_{\rm  marg}(\bm{\lambda})p(\bm{\lambda})$ appearing in Eq. (\ref{eq:post}):
\begin{eqnarray}
\label{eq:EvidenceForEOS}
I(\gamma) = \int dm_1 dm_2 d\chi_1 d\chi_2 p(\chi_1,\chi_2) dz p(z|m_1,m_2) {\cal L}_{marg}(m_1(1+z),m_2(1+z), \lambda_1(m_1|\gamma),\lambda_2(m_2|\gamma),\chi_1,\chi_2)
\end{eqnarray}
where $p(z)$ is the posterior redshift  distribution given a source with masses $m_1,m_2$ was observed, which can be
efficiently extracted from \RIFT{} or tightly constrained by electromagnetic observations.  Appendix
\ref{sec:EOSIntegration} describes   techniques to efficiently evaluate this expression for generic observations.  
Just as with ${\cal L}_{marg}$ and intrinsic binary parameters, we can evaluate the marginalized likelihood $I(\gamma)$
on a  grid of equation of state parameters; interpolate; and generate posterior distributions over those equation of
state parameters, for one or more events.   

\end{widetext}

\begin{figure}
\includegraphics[width=0.45\textwidth]{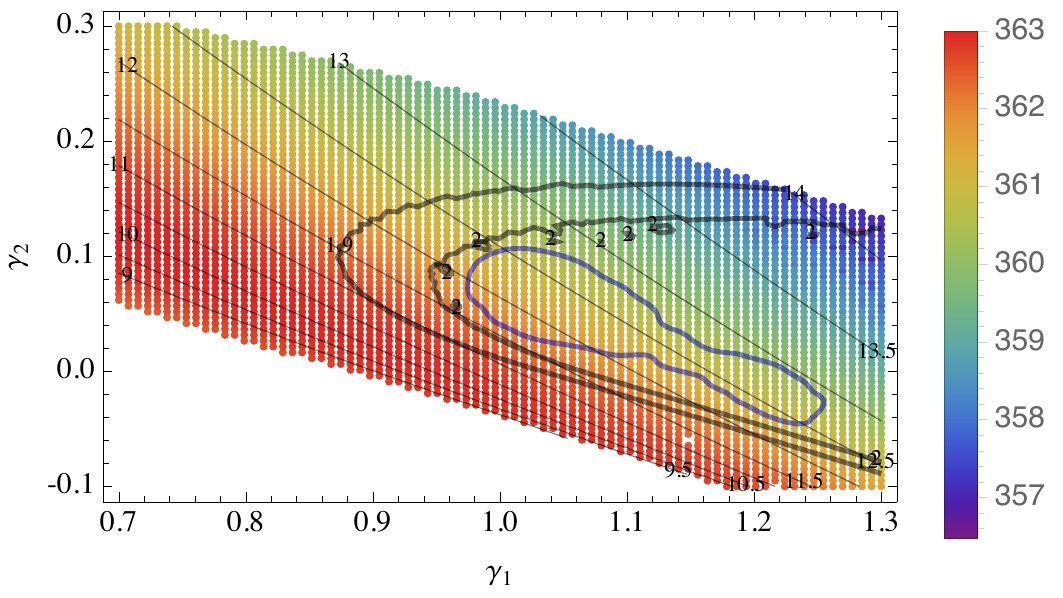}
\caption{\label{fig:EOS}\textbf{Inferences about the equation of state}:
Colors indicate the  marginalized likelihood $I(\gamma)$ versus the two parameters $\gamma_1,\gamma_2$ of the Lindblom et
al spectral EOS representation, applied to a binary neutron star source similar to Tidal-1.  For context, the thin gray lines show contours of
constant radius of a 1.5 $M_\odot$ neutron star, according to this equation of state [$R(m=1.5 M_\odot)$]; the two heavy
black lines show contours of constant maximum mass for $2 M)_\odot$ and $1.9 M_\odot$, respectively.   For the
purposes of this proof-of-concept calculation, we explore only these two variables, fixing the remaining spectral EOS parameters  to $\gamma_3=\gamma_4=0$, $p_o=2.272\times 10^{33}
\unit{dyne/cm^2}$, $x_{max}=7.25$, and $\epsilon_o/c^2=2.05\times 10^{14}\unit{g/cm^3}$.  
The solid blue line shows the 90\% credible interval on the inferred equation of state, after restricting to a causal
EOS and restricting the maximum mass to be greater than  $1.97 M_\odot$.
}
\end{figure}

Figure \ref{fig:EOS} shows a concrete example of this procedure, applied to the spectral equation of state
parameterization introduced by Lindblom \cite{2010PhRvD..82j3011L,2014PhRvD..89f4003L} and to a synthetic BNS generated
using the  APR4 equation of state \cite{1998PhRvC..58.1804A}, as previously analyzed in Figure \ref{fig:tides1}.  For this low-redshift
source, similar to GW170817, we will assume the redshift is well-determined, eliminating that factor in the integrand.
[Equivalently, for this low-redshift source,  the source-frame and detector-frame masses are sufficiently similar that $\lambda(m_{1,z}|\gamma)\simeq \lambda(m_1|\gamma)$
(i.e., the data can't discriminate between them) so the integral over redshift can be performed once and for all, and
all remaining expressions carried out using redshifted masses; see Appendix \ref{sec:EOSIntegration}.]   As a concrete
proof of concept, this figure shows the marginalized likelihood derived from this single event as a function of two spectral equation of state
parameters, as well as Bayesian inferences about those two equation of state parameters when holding others fixed.  
In this analysis, we include only causal equations of state ($v<c$) and require a maximum
neutron star mass greater than $2M_\odot$, motivated by measurements of PSRs 1614-2230 and J0348+0432
\cite{2010Natur.467.1081D,2013Sci...340..448A}.   
More generally, other observational and theoretical constraints will provide a prior $p(\gamma)$ on the equation of
state parameters.   In terms of this prior, the marginal posterior distribution on equation of state parameters is
proportional to $I(\gamma)p(\gamma)$, and the  marginal posterior distribution for any other expression of interest
follows by either quadrature or use of weighted samples, as before.  As a concrete example, if $\gamma_\alpha$ are
$n$ representative EOS configurations from the marginal posterior distribution and, for each $\alpha$, we generate a
Monte Carlo approximation to  Eq. (\ref{eq:EvidenceForEOS}) and retain the weighted samples
 $(w_{k,\alpha},\boldsymbol{\lambda}_{k,\alpha})$ for
$k=1\ldots N$ needed to evaluate it, then the marginal distribution of
$\LambdaTilde$  [Eq. (\ref{eq:lamtilde})] follows by Monte Carlo integration of the cumulative distribution:
$P(<\LambdaTilde) \simeq (n N)^{-1} \sum_{k,\alpha} w_{k,\alpha}p(\gamma_k)\Theta\left(\LambdaTilde - \LambdaTilde(\boldsymbol{\lambda}_{k,\alpha})\right)$.

By providing  marginalized likelihoods $I(\gamma)$, this approach to EOS inference enables the same powerful embarassingly-parallel and
postprocessing-dominated approach used by \RIFT{} itself.  For example, we can combine inferences from multiple events
by simply multiplying the associated likelihood distributions [$I(\gamma|d_1)I(\gamma|d_2)\ldots$].  The end-user is free to efficiently adopt any prior of interest after the
initial analysis; to assess whether choice of priors limits or dominates their analysis; and to incrementally extend
their parameter space exploration with the minimum necessary computational expense.  This approach is well-suited to rapidly-converging and low-dimensional parameterizations
like the spectral method shown in Figure \ref{fig:EOS}
However, due to the remarkably low cost of all integrals involved and our reliance on precomputed marginalized likelihoods,  this approach can also be applied to assess generic EOS
parameterizations inside standard Markov chain  Monte Carlo algorithms.   Though the above discussion is written as if
equation of state dependence enters only through the tidal deformabilities $\lambda$, this procedure works when the
underlying dynamics and radiation models include additional multipolar couplings.
While similar parameterized equations of state have already been developed and used to infer the equation of state from
individual BNS GW measurements \cite{CarneyWade-2018}, our approach will make the best use of multiple observations, as
we don't need additional infrastructure or approximations to estimate $I(\gamma)$.

For low-redshift sources like GW170817 where $\lambda(m_z)\simeq \lambda(m)$ and hence
where the redshift distribution $p(z|m_1,m_2)$ is not needed, and equivalently for sources with known host galaxy
redshifts, we have also explored an alternative approach to compute
$I(\gamma)$.  In this approach, we 
 evaluate $\lnLmarg$ on a grid of mass and spin choices, using the specified EOS to determine
 $\lambda_i=\lambda(m_i|\gamma)$.  We then perform the usual \RIFT{} approach to estimate $\lnL_{\rm marg}$ and hence to
 compute $\int d\bm{\lambda} {\cal L}_{\rm marg}(\bm{\lambda} ) p(\bm{\lambda} )$, a byproduct of the Monte Carlo
 integration procedures used to generate our posterior distributions.   While this approach can be useful for ranking
 the relativity validity  of a few EOS, its duplicative computations and omission of redshift make it too burdensome for
 long-term use.  

\section{Analysis of real events}
In this section, we demonstrate our method can reproduce the interpretation of LIGO coalescing binary black holes.  
Though we do not  not always include  head-to-head examples with the same approximations, these examples demonstrate our
code  functions well even on real LIGO detector noise.   These examples also demonstrate that a feature omitted
from \RIFT{} but included in LI -- marginalization over uncertain detector calibration -- has no impact on our estimates
of intrinsic source parameters. 
Gravitational wave strain data identical to the inputs employed here are available from the LIGO Open Science Center
\cite{LOSC}.

\subsection{Reproducing the interpretation of a real GW source: GW150914}
In this section, we analyze GW150914 with \RIFT{} and LI using several different approximate waveform models.  
In these comparisons, we employ real LIGO data to analyze GW150914,  estimating  strain noise power spectral densities (PSDs) for both events from data segments near each event, similar to the PSDs
used in \cite{LIGO-O1-BBH}.    
We also employ  different spin  priors, such as the volumetric spin prior [Eq. (\ref{eq:prior:spin:volumetric})], than those employed by Abbott et al. \cite{DiscoveryPaper,PEPaper} in the
original analysis of this event. 
Our analysis also differs from these previously-published results insofar as we do not  marginalize over the
calibration uncertainty of the data.  The effect of calibration uncertainty on the intrinsic parameter posterior
distribution is  small. 
Finally, to be consistent with previous analyses with numerical relativity simulations \cite{NRPaper} and keeping in
mind finite model duration, when using
\NRSur{} we adopt a minimum frequency of $30\unit{Hz}$; however, we adopt
a minimum frequency of $20\unit{Hz}$ for our other analyses.  The choice of lower cutoff frequency has little impact on our comparisons \cite{NRPaper}.

\begin{figure*}
\SkipForEarlyCirculation{
\includegraphics[width=\columnwidth]{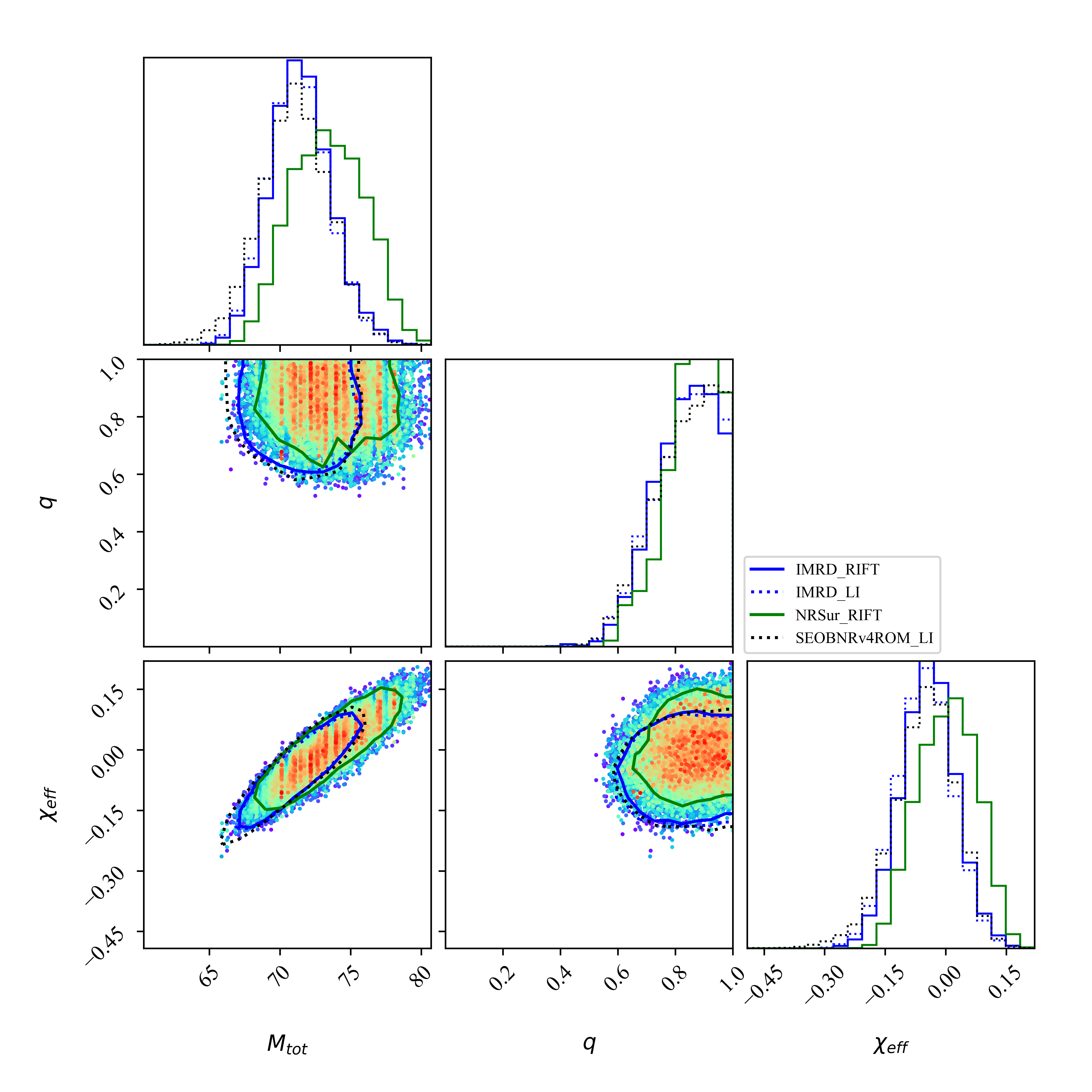}
\includegraphics[width=\columnwidth]{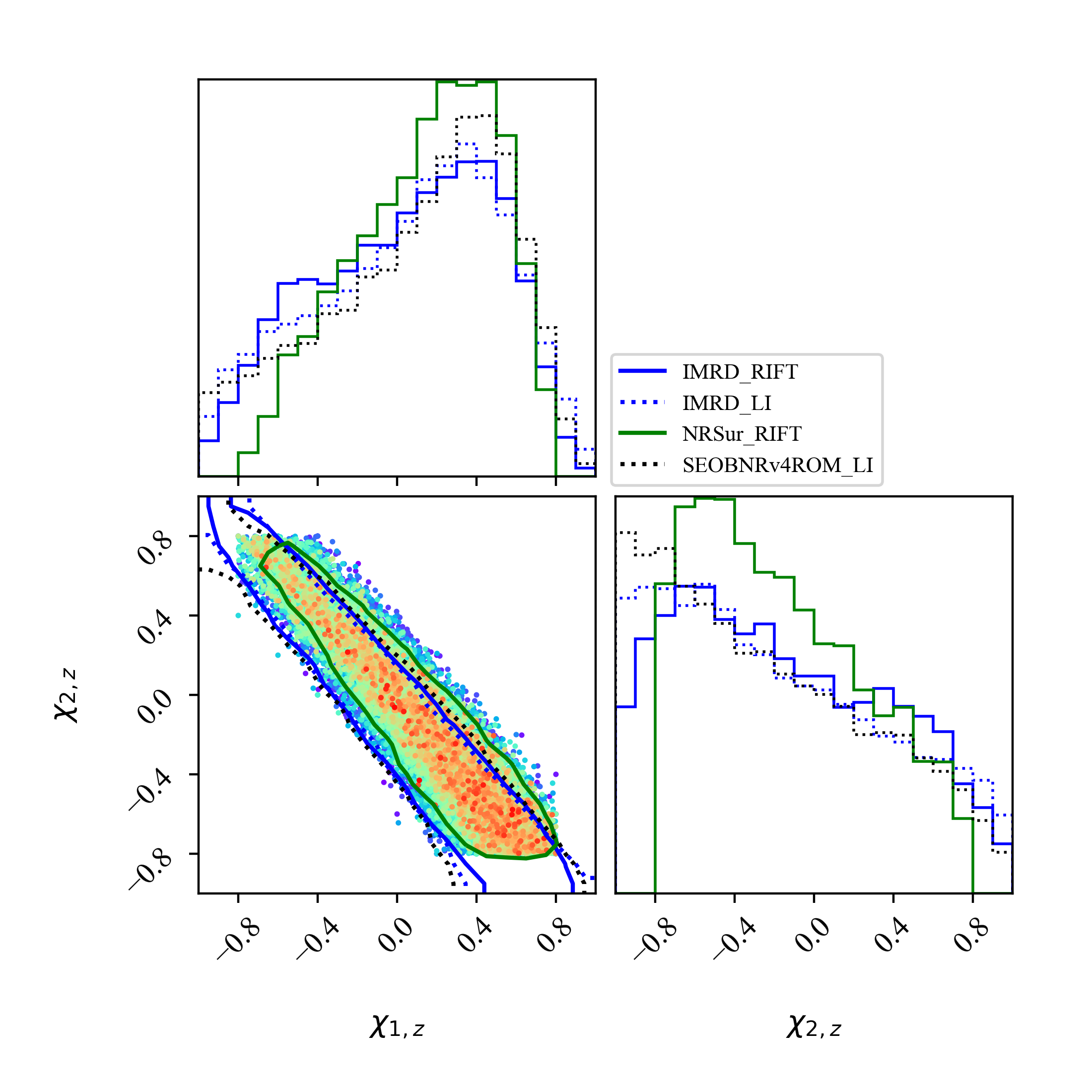}
\includegraphics[width=\columnwidth]{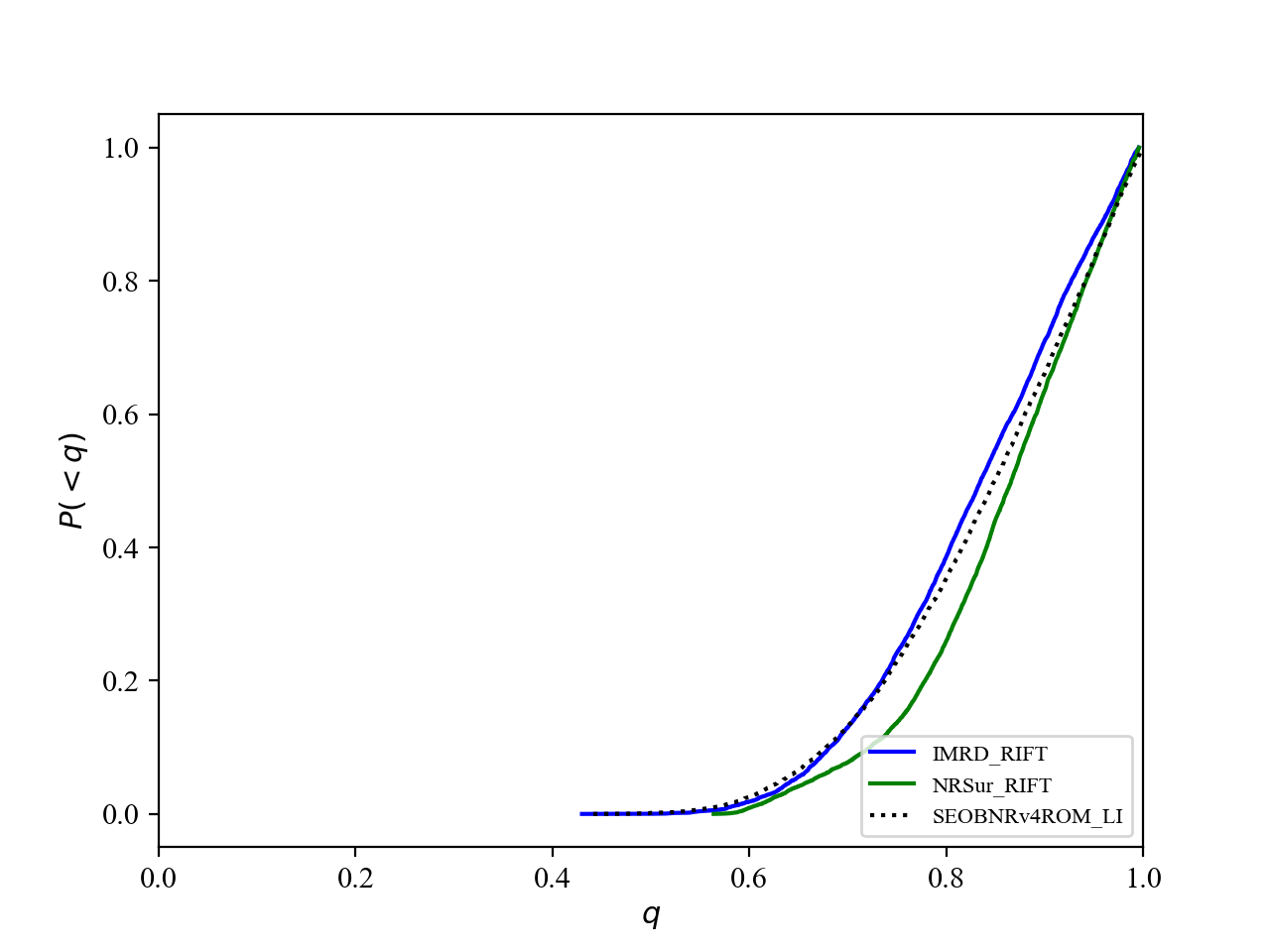}
\includegraphics[width=\columnwidth]{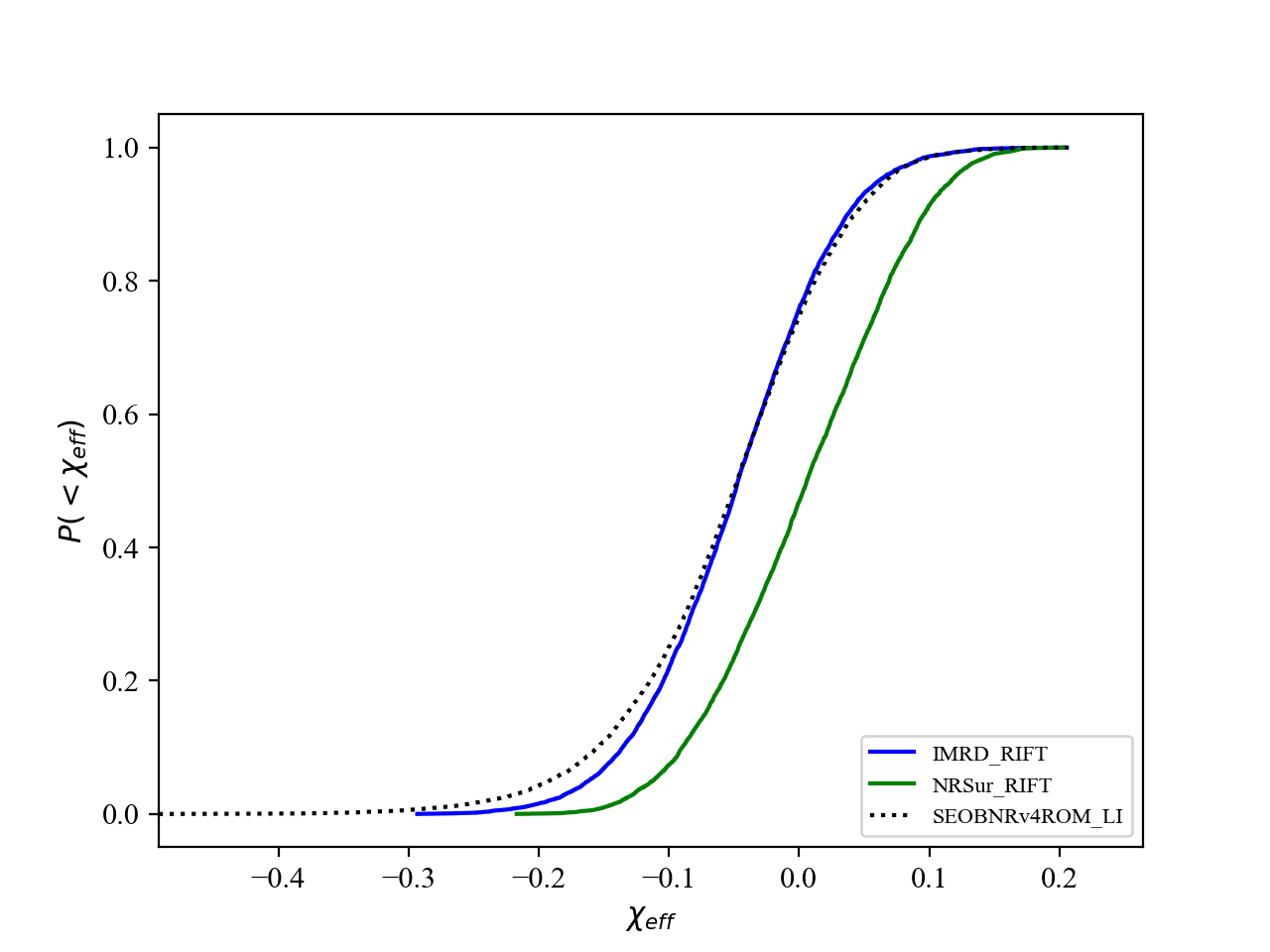}
}
\caption{\label{fig:first1:aligned}\textbf{Reanalysis of GW150914: Aligned models} - This figure shows the results of a reanalysis of the
  first GW detection GW150914, using a prior where each $\chi_{i,z}$ is a uniform random number and the
  assumption that both spins must be parallel to the orbital angular momentum.  The top two
  panels show one- and two-dimensional marginal distributions in redshifted total mass $M_{\rm tot},q,\chi_{\rm eff}$ and
   in the components of each BH's spin $(\chi_{i,z})$ along the orbital angular
   momentum.    The bottom panel provides one-dimensional marginal cumulative distributions for $q$ and
   $\chi_{\rm eff}$, to highlight differences between different approaches.   Curves are colored and styled following
   the convention adopted in previous figures; in particular, black lines correspond to \SEOBAROM{}, solid green lines to
   \NRSur{} with the spins restricted along the orbital angular momentum with all modes up to $\ell_{\rm max}=3$, and dotted blue lines to \IMRPD.  The top right 
  and hence bottom left panels
  show significant differences between \NRSur{} and other approximations principally due to the range of validity of
  the \NRSur{} model, which is valid only for $1/q>0.5$ and $|\chi_i|<0.8$.
}
\end{figure*}

Figure \ref{fig:first1:aligned} shows our analyses of  GW150914 with LI and \RIFT{} using nonprecessing waveform models,
all computed using a prior where  that each $\chi_{i,z}$ is a uniform random number.  
First and foremost, the good agreement between the solid and dotted blue lines (\IMRPD)   demonstrates  how LI and \RIFT{} agree when applied to the same data with the same
noise model, even in real detector noise.  
Second, as in previous result derived by directly comparing GW150914 to numerical solutions of Einstein's
equations \cite{NRPaper}, we find that our posterior inferences derived a waveform model that includes higher modes
(here, \NRSur{} with $\ell\le 3$) more sharply constrains some parameters like the binary mass ratio $q$.   Additionally, as we adopt a more
flexible likelihood model, we may also better constrain several other parameters relative to the analysis presented in
\NRSur{}.  However, our analysis and previous results \cite{NRPaper} suggest that, in an aligned-spin analysis of this
event which adopts a prior not strongly disfavoring anti-aligned spins, the posterior has substantial
support outside of the domain of validity of \NRSur{}.  As demonstrated previously in Figure \ref{fig:mixed1}, the
domain of validity of \NRSur{} can  impact inferences, if the posterior  support extends to the edge of the model
domain.  
In a companion study devoted to direct comparison to numerical relativity solutions,  we will more carefully address
differences between this result and the analysis presented in \cite{NRPaper}, using numerical relativity simulations to
fill the gaps.

\begin{figure*}
\SkipForEarlyCirculation{
\includegraphics[width=\columnwidth]{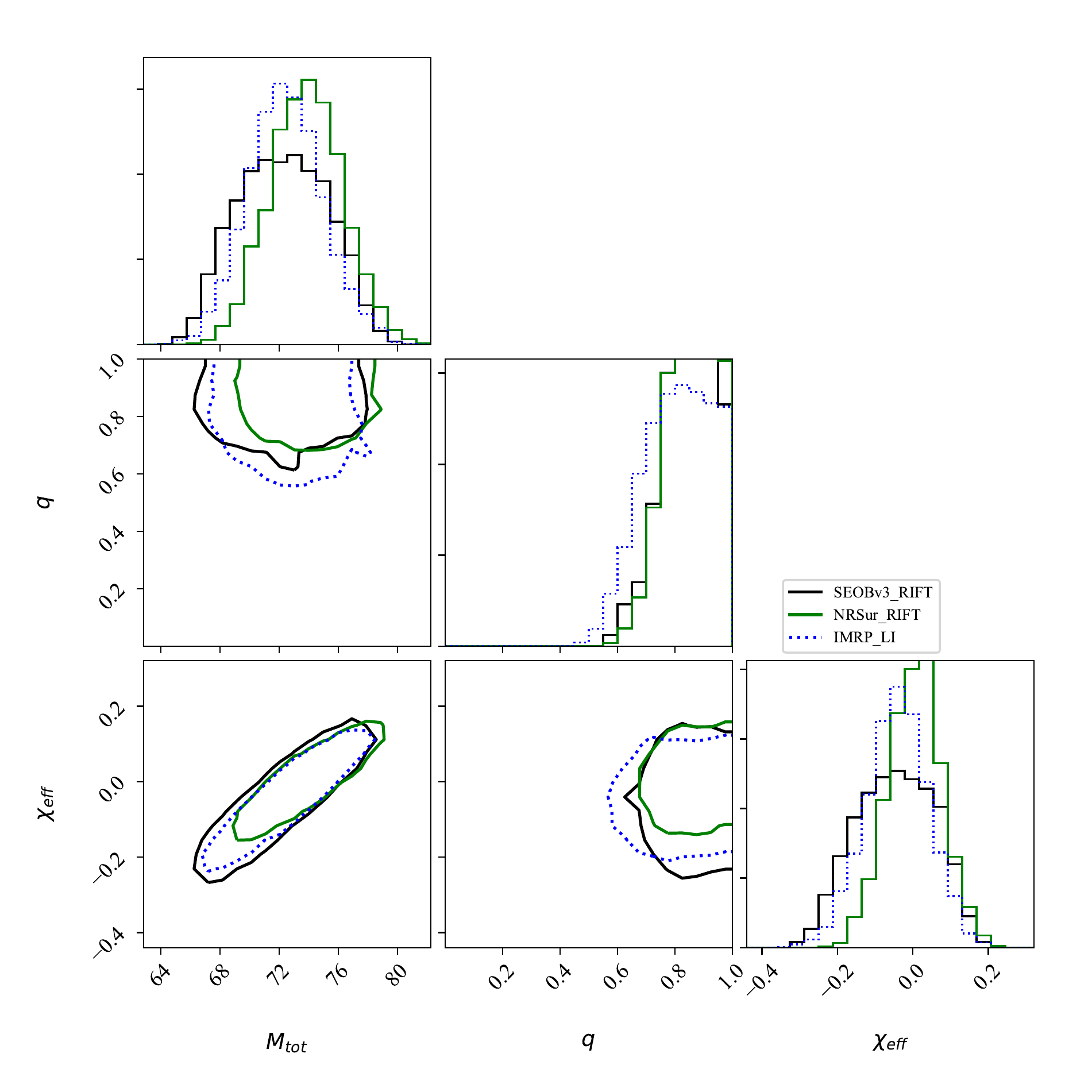}
\includegraphics[width=\columnwidth]{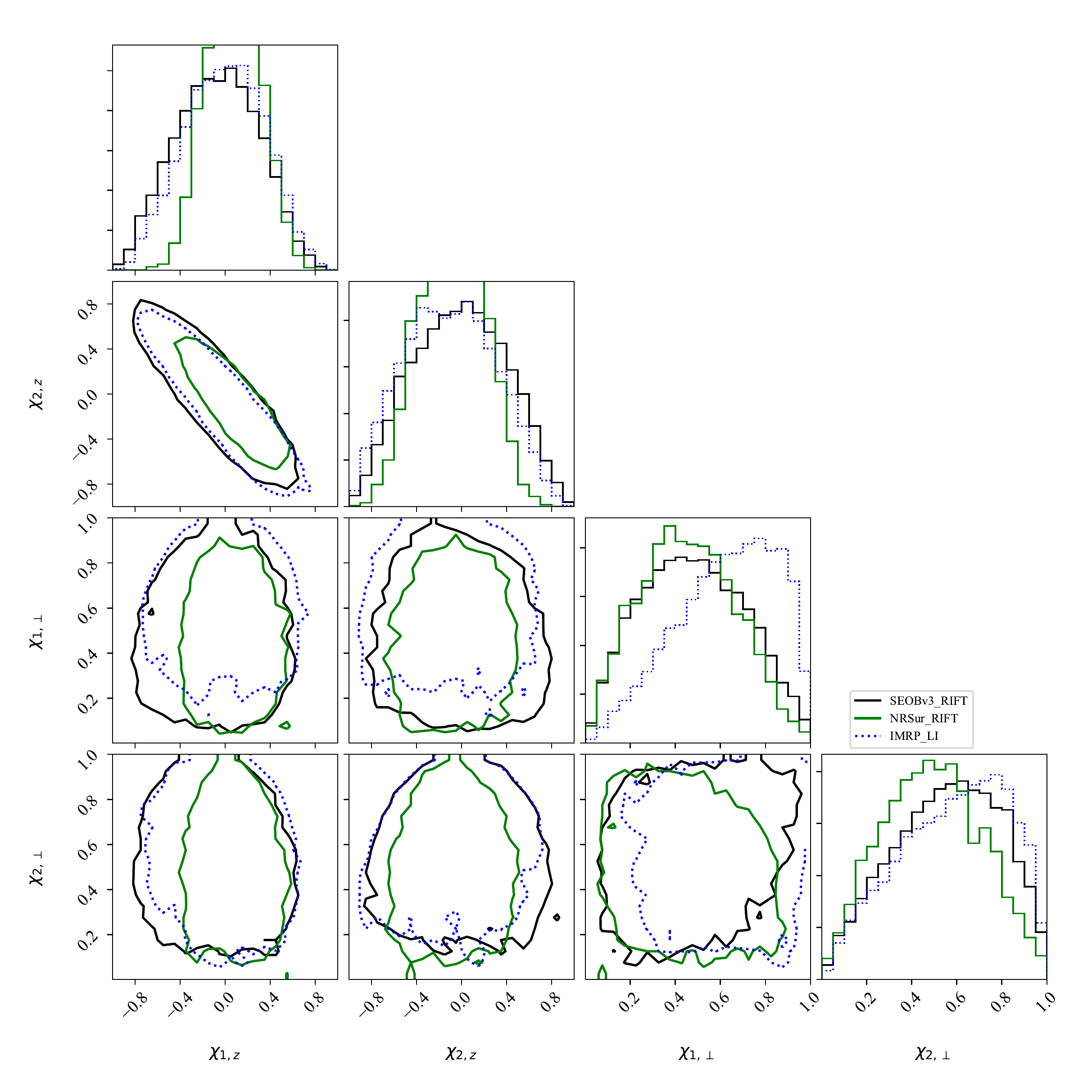}
\includegraphics[width=\columnwidth]{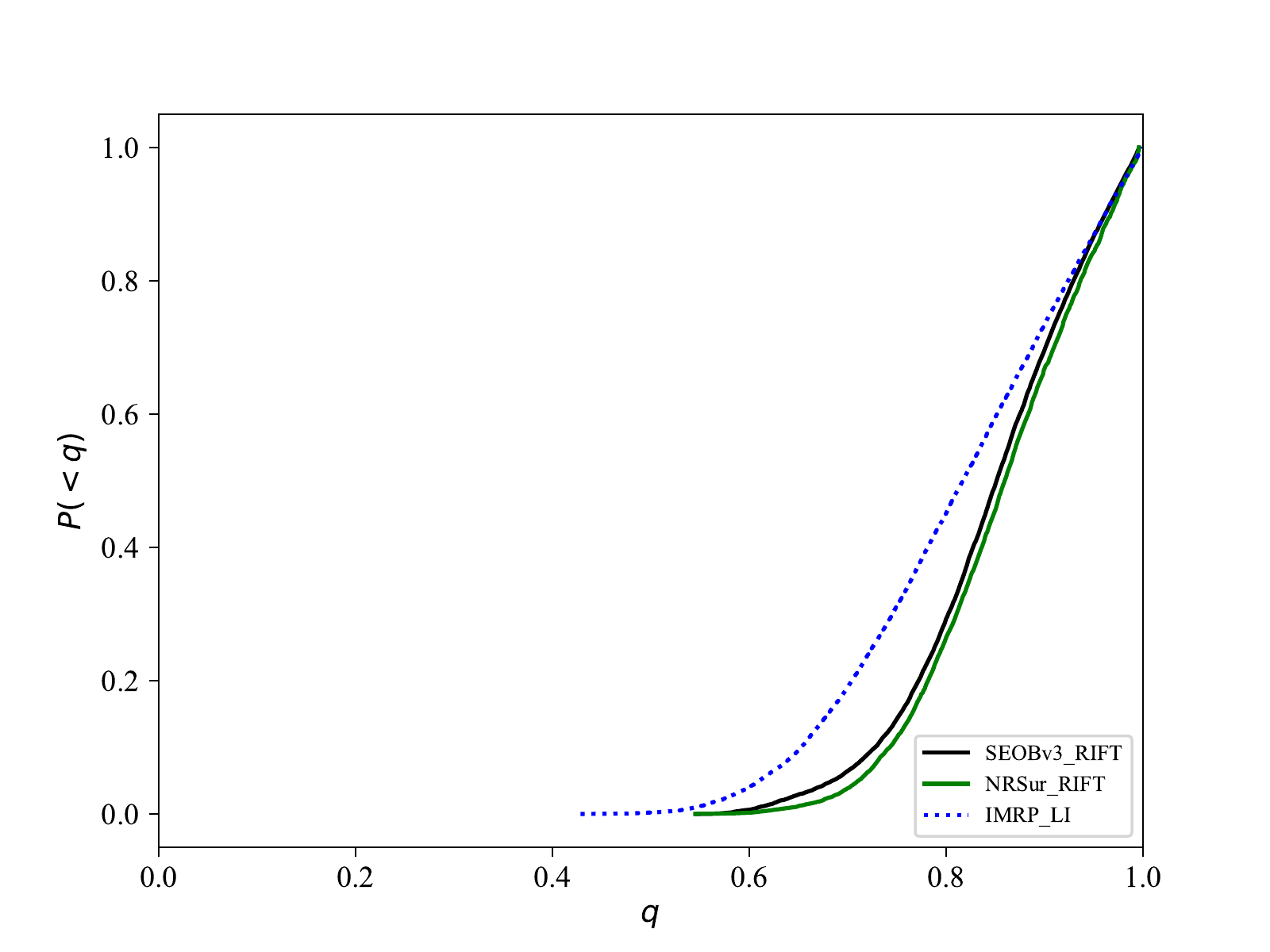}
\includegraphics[width=\columnwidth]{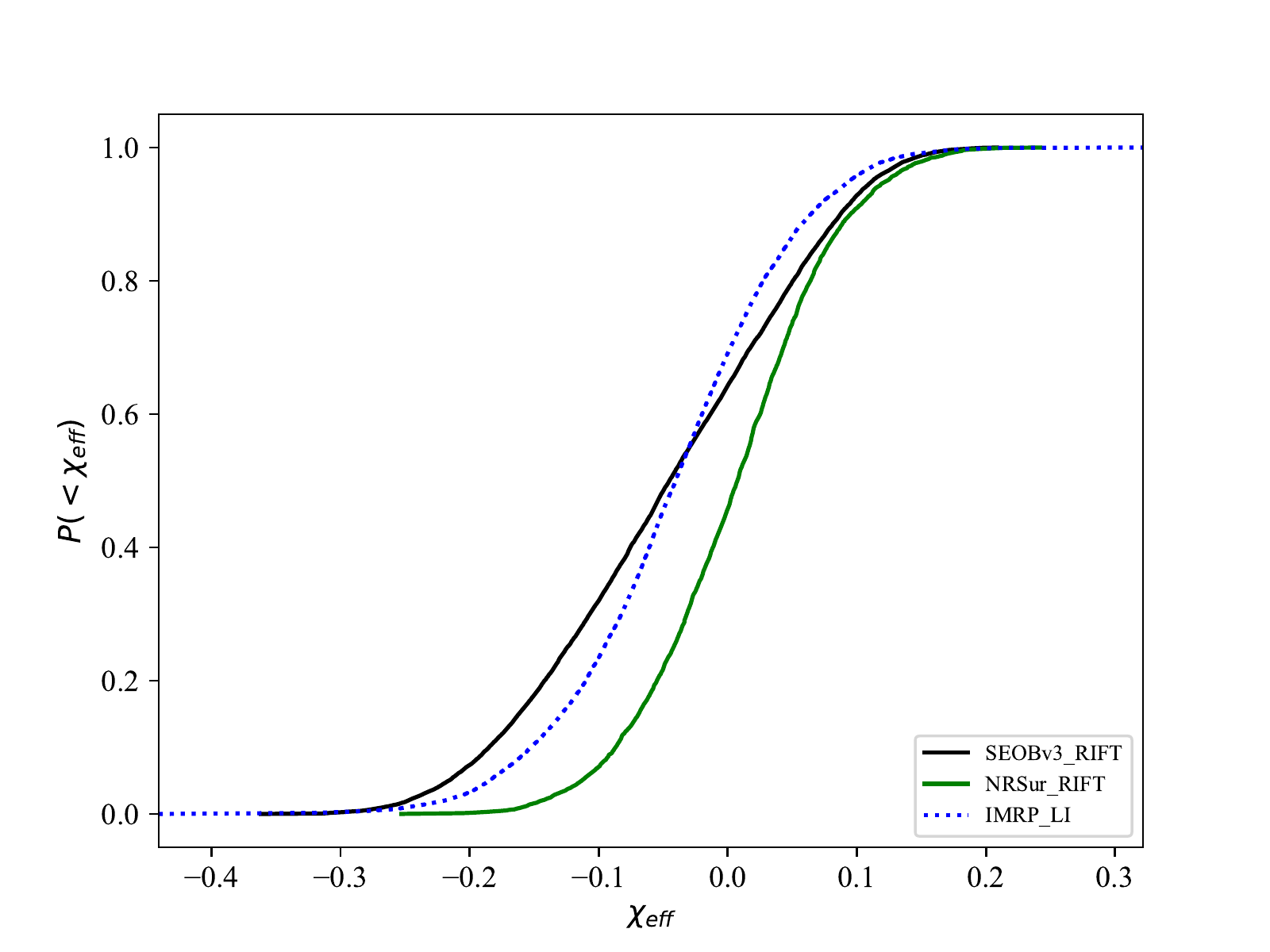}
}
\caption{\label{fig:first1}\textbf{Reanalysis of GW150914: Precessing models} - This figure shows the results of a reanalysis of the
  first GW detection GW150914, using the volumetric spin prior [Eq. (\ref{eq:prior:spin:volumetric})] and waveform
  models that allow for spin-orbit misalignment.  The top two
  panels show one- and two-dimensional marginal distributions in redshifted total mass $M_{\rm tot},q,\chi_{\rm eff}$ and
   in the components of each BH's spin along $(\chi_{i,z})$ and perpendicular $(\chi_{i,\perp})$ to the orbital angular
   momentum.    The bottom panel provides one-dimensional marginal cumulative distributions for $q$ and
   $\chi_{\rm eff}$, to highlight differences between different approaches.   Curves are colored and styled following
   the convention adopted in previous figures; in particular, black lines correspond to \SEOBP{}, solid green lines to
   \NRSur{} with
  fully precessing spins with all modes up to $\ell_{\rm max}=3$, and dotted blue lines to \IMRP.  The top right panel
  shows significant differences between \NRSur{} and other approximations principally due to the range of validity of
  the \NRSur{} model, which is valid only for $1/q>0.5$ and $|\chi_i|<0.8$.
}
\end{figure*}

Next, to provide a comparison that allows for all spin degrees of freedom, we   analyze  with GW150914  \RIFT{}  and \SEOBP  \cite{2016PhRvX...6d1014A} or \NRSur{} ($\ell \le 3$), on the one
hand, and with LI and  \IMRP{}  \cite{DiscoveryPaper,LIGO-O1-BBH} on the other.  
Figure \ref{fig:first1} %
 shows the 90\% confidence intervals for the \RIFT{}+\SEOBP{}, \RIFT{}+\NRSur{} ($\ell \le 3$),  and LI+\IMRP{} analyses
 in black, green, and dotted blue,
 respectively. 
First and foremost,  consistent with our and other prior work, we see differences between the posterior inferences
derived using different waveform approximations and assumptions, even for sources like GW150914 with now-fiducial binary
parameters and source amplitudes.  
Second, consistent with an earlier investigation of GW150914 which directly compared it to  numerical relativity
simulations \cite{NRPaper}, we see that
including higher harmonics enables us to  draw sharper conclusions about binary parameters -- here, the binary mass
ratio.   Our analysis differs from that prior investigation work in two key ways: precessing inference and $\chi$
limits.  While the previous study \cite{NRPaper} compared GW150914 to generic simulations, a posterior distribution was
estimated only on the basis of \emph{nonprecessing} simulations; by contrast, our analysis employs a fully precessing
model.  Conversely, our use of \NRSur{} is limited to $|\chi_i|<0.8$, which directly constraints our ability to draw
generic inferences about BH spins.

\subsection{Reanalysis of LVT151012}

\begin{figure}
\includegraphics[width=\columnwidth]{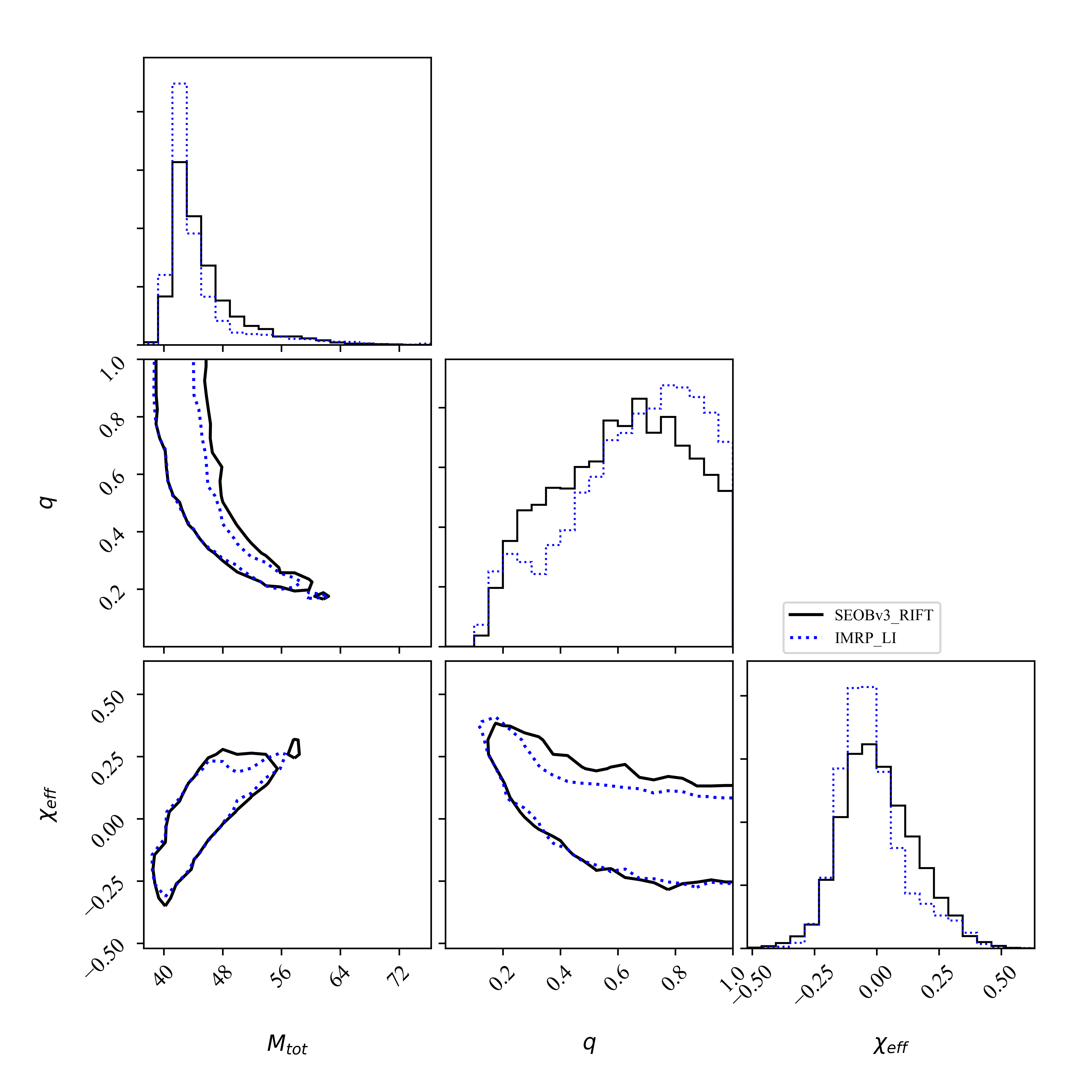}
\caption{\label{fig:LVT151012}\textbf{Reanalysis of LVT151012}  This figure shows two analyses of
LVT151012. The 2D plot shows the PE results in the effective spin ($\chi_{\rm eff}$)-mass ratio (q)
  parameter space (see Eq. \ref{eq:chieff} and Eq. \ref{eq:q} respectively). The other two panels show the corresponding
  1D distributions for each parameter.  Curves and 90\% contours have been colored and styled following the conventions
  of previous analyses.
}
\end{figure}

To demonstrate \RIFT{}'s ability to estimate nongaussian posteriors of low-significance events in real
detector noise, Figure \ref{fig:LVT151012}  presents parameter inferences for LVT151012 with  RIFT{} and LI, performed with
\SEOBP{} (black) and \IMRP{} (blue dotted) respectively.  The analysis uses frequencies between $f_{\rm min}=20\unit{Hz}$ and
$1700\unit{Hz}$, with a volumetric spin prior.   The two analyses demonstrate a level of agreement consistent with previously-demonstrated model systematics.

While we have also performed a corresponding analysis of LVT151012 with \NRSur, including higher harmonics.  Within the
range of mass ratios accessible to \NRSur{}, we see good agreement between an analysis with \SEOBA{} and \NRSur.   However,
due to the limited $q$ domain of \NRSur, we defer a detailed quantitative discussion of higher modes in LVT151012 to
companion work employing numerical relativity simulations to fill the gaps.

\section{Discussion}
\label{sec:conclude}

We have introduced and validated \RIFT{}, a strategy to iteratively produce high-precision posterior distributions of
binary parameters for a wide variety of candidate compact binary coalescences.  We demonstrated by example how this method could employ costly and even heterogeneous
approximations.  As concrete illustrations of its utility, we have employed computationally taxing models, which require
up to one hour per waveform evaluation, to infer parameters of synthetic binary neutron star systems; 
 demonstrated this approach reproduces the
results of other inference strategies, when employing the same approximations; and provided proof-of-concept
demonstrations of new strategies to identify,  assess, and remedy systematic errors which our method uniquely enables.  %

While \RIFT{} does not have the very mature feature set provided by LI, including marginalization over calibration error,
this approach is compatible with these extensions, and can exist simultaneously with other approaches within future parameter inference strategies
for gravitational wave sources.  

The tools demonstrated in this work are actively being generalized.
For example, the gaussian process interpolation strategy can and has been  trivially applied to the (fixed) grid
provided by numerical relativity simulations,  generalizing prior work \cite{NRPaper,LIGO-O1-PENR-Systematics}.   Described in a companion work
\cite{gwastro-PENR-FitHybridPlacement}, this approach was performed to supplement the analysis of GW170104, GW170608,
and GW170814 \cite{2017PhRvL.118v1101A,LIGO-GW170608,LIGO-GW170814}.
Our strategy can also be trivially employed  to construct posterior distributions
that mix approximations, using different approaches in different parts of parameter space, with the boundaries
differentiating between approximations chosen post-facto and with essentially negligible additional computational cost,
besides post-processing.  

While \RIFT{} produced results with dramatically smaller wall-clock time, the investigations performed in this work
 had overall CPU costs comparable to or even in excess of a  corresponding LI analysis with low-cost waveform models.
In the future, when overall computational efficiency becomes a more important constraint on our investigations, we will
describe optimizations to this strategy which reduce our overall computational burden.

\begin{acknowledgements}
The authors thank Prayush Kumar, Juan Calderon-Bustillo,  Chris Pankow, Les Wade, and Matt Carney for constructive discussions during this
investigation, and Ben Farr for feedback on the draft.  
The authors particularly thank Sebastiano Bernuzzi and Alessandro Nagar for their assistance in interfacing with the
latest version of \Resum{}.  
ROS and JAL gratefully acknowledge NSF award PHY-1707965.%
MR gratefully acknowledges support from the RIT College of Science.  
The authors thank to the LIGO Scientific Collaboration for access to the data and gratefully acknowledge the support of
the United States National Science Foundation (NSF) for the construction and operation of the LIGO Laboratory and
Advanced LIGO as well as the Science and Technology Facilities Council (STFC) of the United Kingdom, and the
Max-Planck-Society (MPS) for support of the construction of Advanced LIGO. Additional support for Advanced LIGO was
provided by the Australian Research Council.
Finally, the authors are grateful for computational resources used for the parameter estimation runs provided by Cardiff
University in Cardiff, UK,
funded by an STFC grant supporting UK Involvement in the Operation of Advanced
LIGO; the Albert Einstein Institute at Hanover, Germany; and the LIGO Lab computing facilities at Caltech, Hanford, and
Livingston, maintained by the California Institute of Technology at Pasadena, California.
\end{acknowledgements}
\appendix
\section{Coordinate systems and priors }
\label{ap:CoordinatesAndPriors}

In this appendix, we summarize the coordinate systems and priors adopted to perform the fit and in particular carry out
the Monte Carlo integration procedure for $\lnLmarg$ which produces weighted posterior samples, as described in Section
\ref{sec:Methods}.   The design structure of the  adaptive Monte Carlo integrator developed for \ILE{} and re-used in
our code drives several prior and coordinate choices.  At a very low level, the adaptive procedure assumes the initial true prior $p$
and (adapted) sampling prior $p_s$ are both separable:  $p(x_1\ldots x_n) =p(x_1)p(x_2)\ldots p(x_n)$ for $\{x_1\ldots x_n\}$  the 
coordinate system used to perform the Monte Carlo.   Moreover, because the adaptive procedure optimizes each sampling
prior $p_s$  in one
coordinate at a time, the overall algorithm's performance dramatically improves when the posterior is compatible with
this coordinate system (i.e., the posterior approximately fills a sub-hypercube of our computational domain).  
Whenever we have confident reason to suspect a particular parameter does not impact the posterior distribution
whatsoever, we can analytically marginalize over
 superfluous parameters and perform a fit which omits these parameters.   
Finally, we can explore alternative priors by using weighted  posterior samples, using the ratio of new prior relative
to reference prior in all coordinate dimensions.

\subsection{Masses}
We adopt a uniform prior over the (detector-frame) masses $m_1,m_2$ (with $m_1\ge m_2$), which can be expressed in
several equivalent coordinate systems by a suitable Jacobian transformation.  Following past practice, we nominally
employ a triangular region with $m_i\ge m_{min}$ and $M=m_1+m_2 \le M_{tot,max}$; the prior is $p(m_1,m_2) =
2/(M_{max}-m_{min})^2$, if both regions $m_1>m_2$ and $m_1<m_2$ are included in the integration, and twice that
otherwise.  In most cases, we perform
our underlying calculations in 
$\mc,\eta$ coordinates, where $\mc=(m_1 m_2)^{3/5}/(m_1+m_2)^{1/5}$ and $\eta=m_1 m_2/(m_1+m_2)^2$.  In these
coordinates, the prior can be represented as \cite{gwastro-PE-AlternativeArchitectures}
\begin{align}
\label{eq:prior:McEta}
p(\mc,\eta) = \frac{4}{(M_{max}-m_{min})^2} \frac{\mc}{\eta^{6/5} \sqrt{1-4 \eta}} \; .
\end{align}
As this prior diverges at $\eta=1/4$, it is not well-suited to posterior distributions with significant support very
close to the equal mass line, like binary neutron stars.   For such cases, we change coordinates to $\delta=\sqrt{1-4\eta}$; the prior becomes
\begin{align}
\label{eq:prior:deltaMc}
p(\mc, \delta) = \frac{16 \times  2^{2/5}}{(M_{max}-m_{min})^2} \frac{\mc}{(1-\delta^2)^{6/5}} \; .
\end{align}
In rare cases, we may employ coordinates $M$ and $q=m_2/m_1$; in this case, prior is \cite{2017CQGra..34n4002O}
\begin{align}
p(M,q) = \frac{4}{(M_{max}-m_{min})^2} \frac{M}{(1+q)^2} \; .
\end{align}

For confident detections, the mass posterior will not have significant support outside a compact region.
Because we can freely change the prior in regions far outside the posterior and not impact results, we often adopt
technically convenient boundaries.  For example,  rather than transform the triangular region of permissible masses
$m_1,m_2$ to a more complicated boundary in $\mc,\eta$, we
simply adopt suitable intervals in $\mc$ and $\eta$ which contain the posterior.

\subsection{Spins}
\label{subap:spins}
The spin degrees of freedom are characterized by the spin angular momenta $\mathbf{S}_i$ or equivalently their
dimensionless spins $\mathbf{\chi}_i/m_i^2$.  
For most astrophysically relevant scenarios, the posterior distribution strongly constrains $\chi_{\rm eff}$
[Eq. (\ref{eq:chieff})] but not the complementary combination of aligned spins.
Lacking a compelling astrophysically-motivated choice, several spin-dependent priors have been adopted, which employ
different numbers of spin degrees of freedom.   

\noindent \emph{Volumetric spin prior (generic, cartesian)}: By default, we employ a prior which is uniform in each
component of the dimensionless spins $\mathbf{\chi}_i$ in some hypercube: $\chi_{i} \in [-\chi_{max},\chi_{max}]$.
After generating posterior samples, we then eliminate samples with either dimensionless spin above our threshold
($|\mathbf{\chi}|>\chi_{max}$).   Combined, our hypercube-plus-cut procedure produces an effective prior that is uniform
over the volume of a coordinate sphere of radius $\chi_{\rm max}$:
\begin{eqnarray}
\label{eq:prior:spin:volumetric}
p_{vol}(\chi_i,\theta_{i},\phi_i) = \frac{3}{4\pi \chi_{max}^3} \chi_i^2 \sin \theta_i
\end{eqnarray}
where $\chi_i,\theta_i,\phi_i$ are the spin vector's polar coordinates relative to the orbital angular momentum vector
\cite{2014PhRvD..90b4018F}.  

\noindent \emph{Beta function spin prior (precessing)}: More generally, we can adopt a Beta-distribution prior on the spin magnitude
\begin{align}
p_{\beta}(\chi_i,\theta_i,\phi_i|p,q) = \frac{\Gamma(2+p+q)}{\Gamma(1+a)\Gamma(1+b)}\frac{\chi_i^{p}(\chi_{max}-\chi_i)^{q}}{4\pi \chi_{max}^{p+q+1}}  \sin \theta_i
\end{align}
In this representation, the volumetric spin prior corresponds to $p=2,q=0$.  
The default prior adopted by LIGO when inferring BH spins
uses a uniform spin magnitude distribution
\cite{gw-astro-PE-lalinference-v1}, corresponding to $p=q=0$:
\begin{eqnarray}
\label{eq:pUniformSpin}
p_{umag}(\chi_i,\theta_{i},\phi_i) = \frac{1}{4\pi \chi_{max}} \sin \theta_i
\end{eqnarray}

\noindent \emph{Aligned spin}: When both spins must be parallel to the orbital angular momentum, by default we employ a
uniform prior on $\chi_{i,z}\in [-\chi_{max},\chi_{max}]$.  
To enable comparison to precessing LIGO results, we also employ an alternative prior (``zprior''), proposed by Veitch \cite{VeitchVolumetric2017}
\begin{eqnarray}
\label{eq:zprior}
p_{zprior}(\chi_{i,z}) = \frac{1}{2\chi_{max}} (-\ln |\chi_{i,z}/\chi_{max}|)
\end{eqnarray}
which is equivalent to the uniform spin magnitude prior after marginalizing out other degrees of freedom.  
To demonstrate that $p_{zprior}$ is the corresponding marginal distribution of Eq. (\ref{eq:pUniformSpin}), we perform
  the marginal integral, for convenience denoting spin components for convenience by $x,y,z$ and $\chi_{\rm max}$ by
  $R$.  The cumulative distribution $P(<z)$ can be evaluated
by the following expression when $z<0$:
\begin{align}
P(<z) &= \int_{<z} \frac{dr}{R} \frac{d\phi d\cos \theta}{4\pi}  \\
&= \frac{1}{2R}\int_{-1}^{|z|/R} d\cos \theta \int_{|z|/|\cos \theta|}^R d\cos \theta \\
& = \frac{1}{2R}[z + R - z \ln |z/R|]
\end{align}
and the result follows by differentiation.

\noindent \emph{Correlated, separable spin priors}: In almost all cases, the posterior distribution tightly constrains
$\chi_{\rm eff}$ [Eq. (\ref{eq:chieff})]
but leaves the complementary degree of freedom almost completely unconstrained.  To accelerate sampling in the common
case with $m_1\simeq m_2$, we can adopt a uniform spin prior in $\chi_{z,\pm} = (\chi_{1,z}\pm \chi_{2,z})/2$, then
eliminate samples which have $\chi_{z,i}$ otherwise inconsistent with the limit imposed by $\chi_{max}$.  This approach
can be directly employed to accelerate sampling with both the aligned uniform spin prior and the precessing volumetric
prior; results for non-separable priors follow by reweighting posterior samples.

\noindent \emph{Marginal prior for $\chi_{\rm eff}$ (uniform)}:  Because the individual spin components $\chi_{i,z}$ are
rarely observationally accessible, one can imagine marginalizing out the superfluous
degree of freedom, reducing the marginal likelihood and related Monte Carlo calculations to integrals like 
$I = \int_{|\chi_1|,|\chi_2| < 1}
f(\chi_{\rm eff}) d\chi_{1,z}d\chi_{2,z}$ via
\begin{eqnarray}
I = \int f(\chi_{\rm eff}) d\chi_1 d\chi_2 = 4 \int f(\chi_{\rm eff}) p(\chi_{\rm eff}|q) d\chi_{\rm eff}
\end{eqnarray}
where for simplicity we adopt a uniform spin magnitude prior.  
This expression is not directly applicable to our low-level Monte Carlo technique, as this spin prior depends on mass ratio.  Such an
expression will however be a useful reference when we want to rescale posterior samples to alternative prior
distributions.  

Reweighting is only successful if the posterior has broad support.  For nonprecessing inference, while analyses performed with Eq. (\ref{eq:zprior}) generally have support concentrated near to zero
spin, an analysis with the uniform spin magnitude prior will have support generally for all $\chi_{i,z}$ consistent with
the likelihood.   When constructing a fiducial  marginal $\chi_{\rm eff}$ distribution, we therefore derive it under the
assumption of uniformly distributed $\chi_{i,z}\in [-\chi_{max},\chi_{max}]$.  In the interests of clarity and without
loss of generality -- all results scale linearly with $\chi_{max}$ -- in the derivation that follows we adopt
$\chi_{max}=1$. %

To evaluate $p(\chi_{\rm eff}|q)$, we first define a helpful shorthand (to avoid ambiguity)
\begin{eqnarray}
g(a,b) = \frac{m_1 a + m_2 b}{m_1+m_2} = (a + q b)/(1+q)
\end{eqnarray}
which is $\chi_{\rm eff}$.  The integrand and prior has four natural breakpoints at $g(\pm 1,\pm 1)$, ordered so
$g(-1,-1)\le g(-1,1)\le g(1,-1) \le g(1,1)$.  Within each region, we can do the integral $\int d\chi_1 d\chi_2
\delta(z-g(\chi_1,\chi_2))$ simply by keeping track of the limits of integration:
\begin{align}
J(z) &= \int_{-1}^1 d\chi_1 \int_{-1}^1 d\chi_2\delta(z-g(\chi_1,\chi_2)) \\
&= \int_{\chi_2 = \chi_2(\chi_1,\chi_{\rm eff}) } d\chi_1 (1+q)/q \\
&= \frac{(1+q)}{q} [\chi_{1,+} - \chi_{1,-}]
\end{align}
where $\chi_{1,\pm}$ are the largest and smallest allowed values of $\chi_1$ for a given choice of $\chi_{\rm eff}$.
Looking at the square, when $\chi_{\rm eff} < g(-1,1)$, we know $\chi_{1,-}=-1$ and when $\chi_{\rm eff}>g(1,-1)$ we
know the upper bound is $1$.  Otherwise, we know $\chi_{1,\pm}$ occurs when $\chi_{2}=\pm  1$, implying
\begin{eqnarray}
\chi_{1,\pm}(z) = \pm q + z(1+q) 
\end{eqnarray}

We find the following expression for our marginal prior
\begin{eqnarray}
p(z|q) = \frac{1+q}{4q} \times \begin{cases}
1 - \chi_{1,-}  & z \in[ g(1,-1), g(1,1)] \\
\chi_{1,+} - \chi_{1,-}  = 2 q  & z \in[ g(-1,1), g(1,-1)]  \\
\chi_{1,+}+1  & z \in[ g(-1,-1), g(-1,1)] 
\end{cases}
\end{eqnarray}

\begin{widetext}
\subsection{Tides}
\label{subap:tides}
The tidal deformability of each compact binary can be characterized by a dimensionless parameter $\Lambda_i$, which is
zero for black holes.   By default, we adopt a uniform prior on $\Lambda_i\in [0,\Lambda_{max}]$.

The leading-order effects of tidal deformation enter into the gravitational wave signal via two quantities
$\widetilde{\Lambda}, \delta\widetilde{\Lambda}$
\begin{align}
\label{eq:lamtilde}
\LambdaTilde &=
\frac{16}{13} \frac{
(m_1+12 m_2)m_1^4\Lambda_1 + (m_2 + 12 m_1) m_2^4\Lambda_2
}{
(m_1+m_2)^5
} \nonumber \\
&=
 \frac{8}{13}\bigg{[} (1+7\eta-31\eta^2) (\Lambda_1 + \Lambda_2) + \sqrt{1-4
    \eta}(1+9\eta-11\eta^2)(\Lambda_1-\Lambda_2) \bigg{]} \\
\DeltaLambdaTilde &= 
\frac{1}{2}\bigg{[} \sqrt{1-4\eta}\bigg{(}1-\frac{13272}{1319}\eta +\frac{8944}{1319}\eta^2\bigg{)}(\Lambda_1+\Lambda_2)+\bigg{(}1-\frac{15910}{1319}\eta+\frac{32850}{1319}\eta^2+\frac{3380}{1319}\eta^3\bigg{)}(\Lambda_1-\Lambda_2)\bigg{]},
\end{align}
In most cases of current astrophysical interest,  $\LambdaTilde$ can be weakly constrained and $\DeltaLambdaTilde$
cannot be constrained at all.    

Because of the constraint that $\Lambda_{i}\ge 0$,  a corner in tidal parameter space, the marginal distribution of $\LambdaTilde,\DeltaLambdaTilde$ near
$\LambdaTilde\simeq 0$ increases linearly with $\LambdaTilde$.  To more transparently reflect the astrophysical
significance of the posterior distribution of $\LambdaTilde$, it is helpful to adopt a prior corresponding to a
\emph{uniform} distribution of $\LambdaTilde$.
Once again, such an alternative prior depends on binary mass ratios and therefore is useful only for post-processing and reweighting, not
as part of our initial Monte Carlo analysis which requires separable priors.  
The underlying calculation of the marginal prior on $\LambdaTilde$ follows exactly like the calculation for $\chi_{\rm
  eff}$ above. 

\ForInternalReference{
\editremark{finish me}
One simple constructive approach to alternative priors
\editremark{insert}
\begin{eqnarray}
\begin{bmatrix}
\tilde{\Lambda} \\
\Delta \tilde{\Lambda}
\end{bmatrix}
=
\begin{bmatrix}
a & b \\
c & d
\end{bmatrix}
\begin{bmatrix}
\Lambda_1 \\
\Lambda_2
\end{bmatrix}
\end{eqnarray}
}

\end{widetext}

\section{Supplementary validation studies of iterative posterior generation}
\label{ap:validate:Gaussians}

In this section we describe systematic tests of the general algorithm and specific implementation described in  Section
\ref{sec:Methods} to reconstruct the posterior distribution by means of iteratively fitting the likelihood distribution,
then drawing candidate points from the posterior distribution. 

\begin{figure}
\includegraphics[width=\columnwidth]{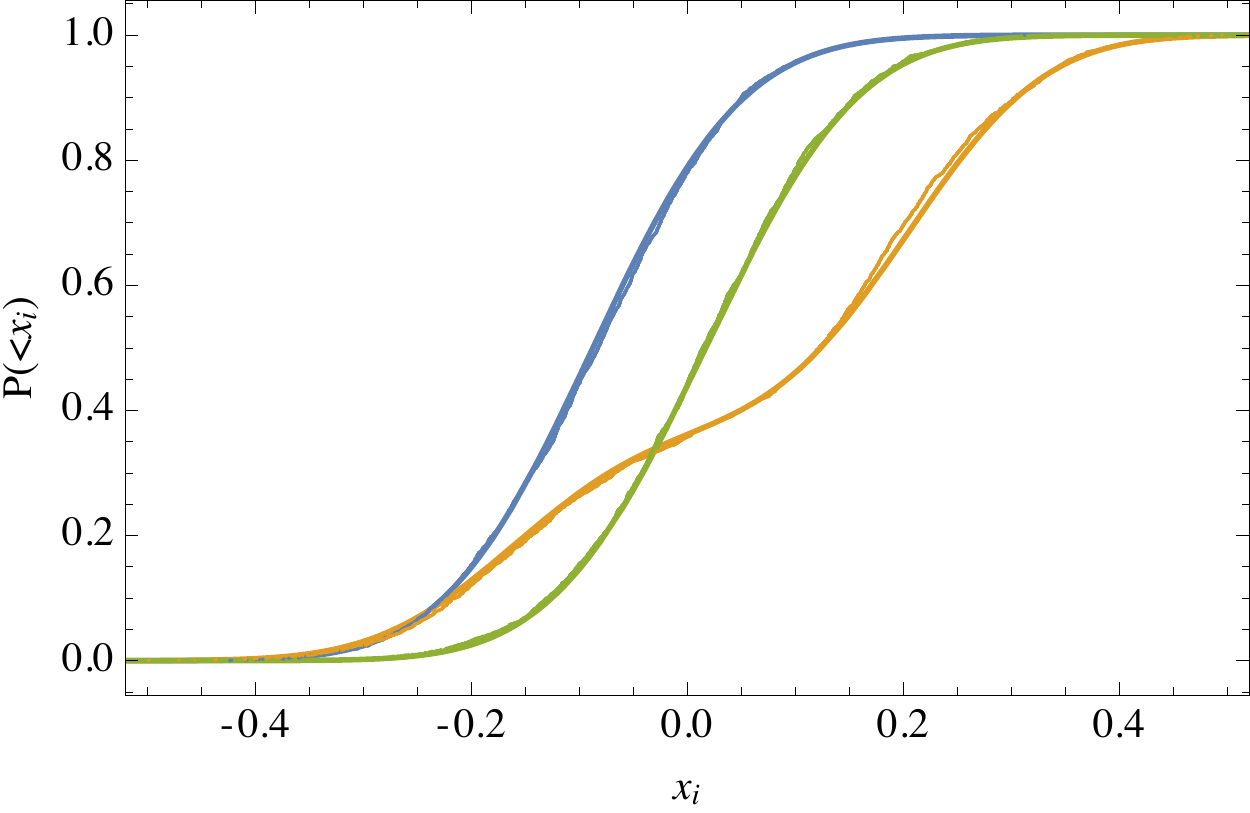}
\includegraphics[width=\columnwidth]{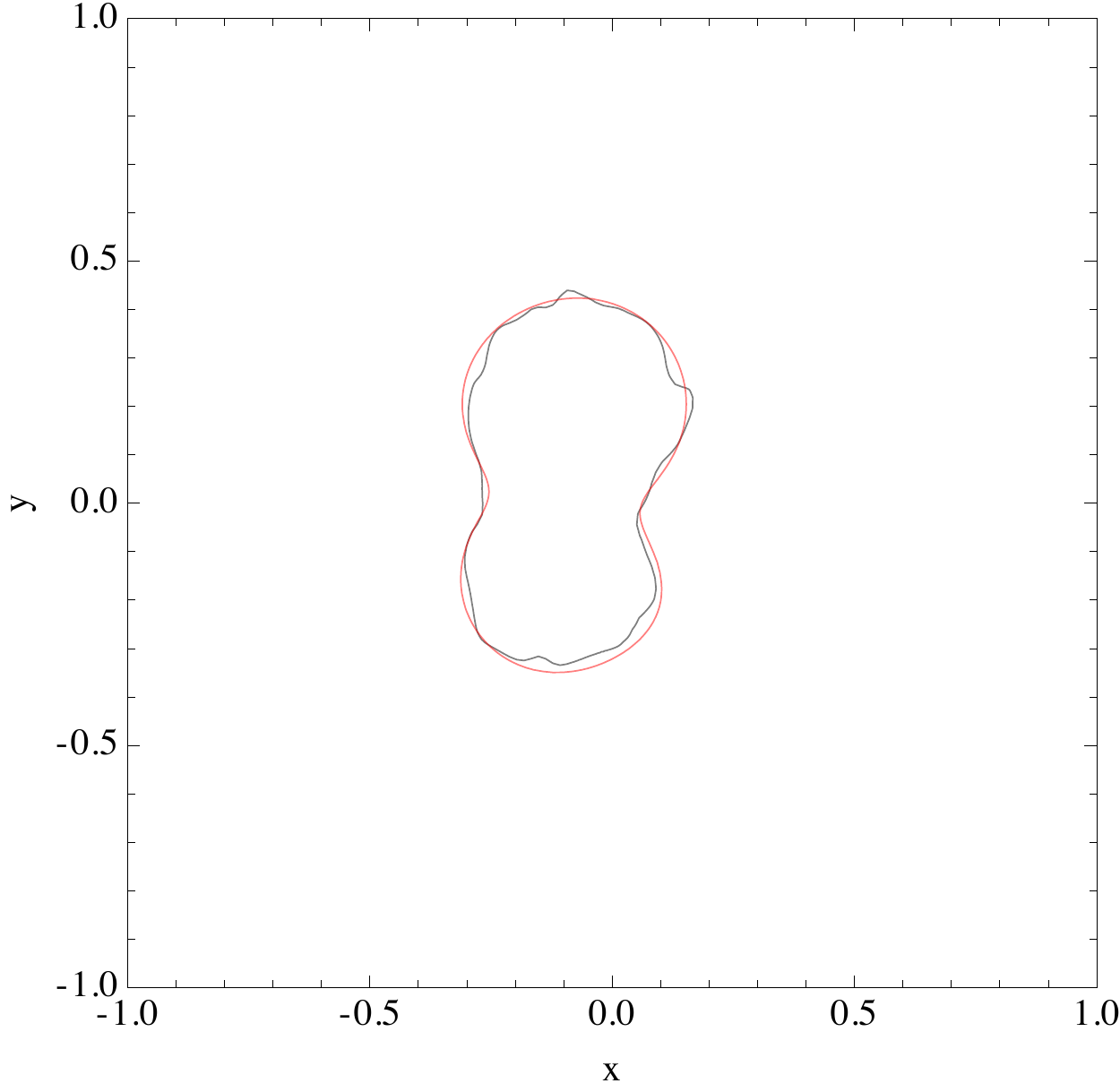}
\caption{\label{fig:IllustrateDiagnostics}\textbf{Reliable reconstruction in three dimensions}: 
Parameter inference of a $3$ dimensional gaussian mixture model with 4 components, compared with the analytic posterior distribution.
\emph{Top panel}:  The \emph{marginal, 1-dimensional} posterior estimates extracted from a mixture of four  randomly-centered $3$-dimensional
  gaussians.  Solid lines show the cumulative 1d distributions for each of the three
  dimensions; points show the estimated distributions from our code. The numbers on the left provide an
  (approximate) KL divergence between each marginal 1d distribution and the known 1d marginal distribution. The first
  number refers to the first (blue) curve, and so forth (yellow, green, $\ldots$).  
\emph{Bottom panel}: The 90\% credible interval derived from the exact (red) and approximate (black) two-dimensional
marginal distribution.   The black curve is estimated based on $2000$ samples drawn from the posterior distribution.
}
\end{figure}

In these controlled tests, we generate synthetic likelihood function on a hypercube $x\in [-1,1]^d$ and a gaussian mixture model:
\begin{eqnarray}
{\cal L}(x) = {\cal L}_{\rm ref} p(x) = {\cal L}_{\rm ref}\sum_k \frac{w_k}{(2\pi \sigma^2)^{d/2}} e^{-(x-\mu_k)^2/2\sigma^2}
\end{eqnarray}
where $w_k$ are weights with $\sum_k w_k=1$. 
We drew random weights; random gaussian centers $\mu_k$ in the hypercube $[-0.6,0.7]/\sqrt{d}$; and for convenience
fixed $\sigma=0.1$.
We drew initial points randomly from the hypercube, seeded by a few points from the true posterior distribution, then
applied our iterative code, iterating five times.   Figure \ref{fig:IllustrateDiagnostics} shows an example of the output of our code,
compared to the analytic one- and two-dimensional posterior distributions for a generic three-dimensional and
four-component gaussian mixture model. 
 To quantify agreement between the two distributions, we also provide an (approximate) KL
divergence $\int dz  p_1(z) \ln p_1(z)/q_1(z)$, where $p_1(z)$ is the exact 1d marginal distribution and $q_1(z)$ is our 
approximate 1d marginal distribution.
A fairly accurate reconstruction will have $D_{KL} \lesssim 2\times 10^{-2}$.   
For  arbitrary gaussian mixture models in dimensions $d\le 6$, we confirmed our approach consistently reproduces one- and two-dimensional posterior marginal distributions, with small error.

\section{Evaluating the evidence for an equation of state}
\label{sec:EOSIntegration}
In Section \ref{sec:sub:EOS}
we described a simple integral over source intrinsic parameters and redshift [Eq. (\ref{eq:EvidenceForEOS})] to assess
the compatibility with a given observation and a proposed equation of state, characterized by hyper-parameters $\gamma$
which determine a relationship between tidal deformability $\lambda$ and source-frame NS gravitational mass $m$:
$\lambda=\lambda(m|\gamma)$.  
Because GW measurements naturally very tightly constrain the redshifted chirp mass $\mc_z=\mc(1+z)$ to a characteristic
range of scale $\sigma_{\mc_z}$, the integrand is nearly
zero except for a very narrow range of  $\mc_z$ centered on $\mc_{z,*}$.  Additionally, to an excellent approximation the integrand does not
depend on $\DeltaLambdaTilde$, and the function $\lambda(m)$ is effectively constant when the mass changes by of order
$\sigma_{\mc_z}$.   We therefore change variables to $\mc_z,\eta,z,\chi_i$ [Eq. (\ref{eq:prior:McEta})], then perform
the integral over $\mc_z$ and $\chi_i$:
\begin{align}
I(\gamma) &= \int dz d\eta \frac{4 p(z|\mc_{z,*}) 
  G\left(\eta,\LambdaTilde(\eta,\mc_{z,*})\right)}{(1+z)^2(M_{max}-m_{min})^2 \eta^{6/5}\sqrt{1-4\eta}}  \\
G(\eta,\LambdaTilde) &= \int d\mc \mc  {\cal L}(\mc,\eta,\LambdaTilde) d\chi_1 d\chi_2 p(\chi_1,\chi_2) 
\end{align}
where $\LambdaTilde_*(\eta,z)$ follows by evaluating $\lambda_i(m_i)$ using the masses $m_i$ derived from $\eta$ and the
appropriate source-frame chirp mass $\mc_{z,*}/(1+z)$, and where $p(z|\mc_{z,*})$ is the fully marginalized distance distribution implied
by this observation for a source with detector-frame chirp mass $\mc_{z,*}$.  Following  Eq. (\ref{eq:prior:deltaMc}),
changing coordinates to $\delta$ rather than $\eta$ removes the integrable singularity at $\eta=1/4$ and makes numerical methods more robust.

\ForInternalReference{

\section{Supplementary illustrations}

In this appendix, we illustrate the behavior of and need for several features of our approach.

Figure \textbf{XXX} compares the distance and source-frame mass distributions recovered by our procedure and LI. \editremark{makeme}

Figure \textbf{XXX} shows several snapshots of the iterative distribution generation process.

Figure \textbf{XXX} illustrates how dithering improves convergence.
}

\bibliography{paperexport,LIGO-publications,textbooks}
\bibliographystyle{unsrt}

\end{document}